\documentclass[twocolumn,aps,superscriptaddress,preprintnumbers]{revtex4-2}

\usepackage{transparent}
\usepackage{amsmath}
\usepackage{graphicx}
\usepackage{amssymb}
\usepackage{epstopdf}
\usepackage{braket}
\usepackage{mathtools}
\usepackage{bm} 
\usepackage{longtable}
\usepackage{multirow}
\usepackage{enumerate}
\usepackage{hyperref}
\usepackage{xcolor}
\usepackage{graphicx} 
\usepackage{physics}
\usepackage{array}
\usepackage{lipsum}
\usepackage{tabularx}
\usepackage{printlen}
\usepackage{textcomp}

\usepackage{blindtext}

\usepackage{array}

\usepackage{sidecap}
\hypersetup{colorlinks=true,       
            linkcolor=blue,        
            citecolor=blue,        
            filecolor=blue,        
            urlcolor=blue,         
            runcolor=blue
}
\usepackage{svg}
\usepackage{tikz,lmodern}
\begin{document}

\title{Ultrafast single-photon interference with a dipole qubit in a nanocavity}

\author{Athul S. Rema}

\author{Adri\'an E. Rubio López}

\affiliation{Department of Physics, Universidad de Santiago de Chile, Av. Victor Jara 3493, Santiago,  9170124, Chile.}

\author{Felipe Herrera}
\affiliation{Department of Physics, Universidad de Santiago de Chile, Av. Victor Jara 3493, Santiago,  9170124, Chile.}
\affiliation{Millennium Institute for Research in Optics, Concepci\'on, Chile.}
\email{felipe.herrera.u@usach.cl}

\date{\today}           

\begin{abstract}
The stationary spectrum of individual dipole emitters in plasmonic nanocavities has been studied for a range of cavity geometries and dipole configurations. Less is known about the coherent dynamics of single photon creation in the nanocavity near field by an excited dipole. We address this gap by developing a Lorentzian kernel approximation that solves the time-dependent Schr\"odinger equation that describes the coupled dipole-photon dynamics in the single-excitation manifold. Our approach encodes the broadband nature of the nanocavity field through a non-Markovian memory kernel, derived from macroscopic QED theory. For a two-level dipole near a metallic nanosphere, we show that the single photon probability density in frequency space evolves in strong coupling from an initially localized source at the qubit frequency into a Rabi doublet over a timescale governed by the kernel spectrum. This dynamical crossover is accompanied by the formation of single-photon interference patterns in frequency and time, propagating coherently over a timescale limited by the shape of kernel spectrum to $\sim 100-150$ fs, which is accessible to ultrafast spectroscopy. We also show that the stationary spectrum of the coupled system can be manipulated by driving the nanocavity field using coherent pulses with variable spectral bandwidth. Using single-photon pulses  narrower than the kernel spectrum, the Rabi splitting in a system that supports strong coupling can be effectively removed. The applicability of our results to other dipole-nanocavity configurations is discussed and a general strong coupling criterion for nanocavities is formulated.
\end{abstract}
\maketitle

\section{\label{sec1}Introduction}

Manipulating the wavefunction of individual photons is a fundamental task in quantum information \cite{Couteau2023}. Schemes are available to modify the temporal, spectral, polarization and spatial properties of individual photons across the electromagnetic spectrum, using atomic gases \cite{Kimble1977,Kuhn2002,McKeever2004}, inorganic semiconductors \cite{Buckley2012}, defects in solids \cite{Schroder2011}, metallic nanostructures \cite{Koenderink2017}, organic molecules \cite{Lettow2010,Toninelli2021} and superconducting circuits \cite{Houck2007,Forn-Diaz2017}. Practical uses of single photons imposes constraints on the rate, purity, indistinguishability, and controllability of individual photon emission events \cite{Aharonovich2016}, which ultimately depend on fundamental light-matter interaction processes that are specific to the emitting material system \cite{Sipe1995}. For applications that require narrowband single photons with high indistinguishability, such as quantum networks \cite{Azuma2023}, atomic or atom-like emitters with narrow energy levels are used \cite{Gheri1998,Scholz2009,Muller2017}. Deterministic control of emission events is achieved using external fields to modify the emitter spectrum \cite{Kolchin2008,Jin2014,Morin2019}. For systems with broad spectrum due to interaction with an environment, such as atoms near metal surfaces \cite{Ganesan1996}, decoherence limits the applicability of known single-photon control schemes. Alternative mechanisms must be developed to measure and eventually manipulate single photon coherence for broadband emitters. 

Few-level dipoles in the near field of metallic nanoparticles are commonly studied broadband quantum emitters \cite{Gonzalez-Tudela2024}. Localized surface plasmon resonances confine and amplify the electromagnetic field at the emitter location with spatial resolution below the diffraction limit \cite{Barnes2003}, leading to Purcell enhancement of spontaneous emission in weak coupling \cite{Anger2006,Pelton2015,Barnes2020} and coherent hybridization of light and matter in strong coupling \cite{Torma2014,Delga2014,Haran2019}. These coupling regimes are universal in cavity quantum electrodynamics (QED) \cite{Tame2013} and have been experimentally studied in plasmonic nanocavities from ultraviolet to infrared frequencies using different types of material dipoles \cite{Wan2016,Baranov2018}. 

The sub-picosecond radiative and non-radiative decay of surface plasmons limits the lifetime of material dipole coherence \cite{Kyle2018}, constraining the temporal and spatial resolution needed to address individual photons produced in plasmonic nanocavities. Conventional diffraction-limited techniques to measure and control propagating single photons \cite{Specht2009,Davis2018,Pursley2018,Fenwick2025} should then be supplemented with novel imaging and spectroscopy techniques that offer nanometer and femtosecond spatiotemporal resolution \cite{Hillenbrand2000,Ocelic2006,Zhao2025} to measure quantum optical coherence in nanocavities.  

In this work, we develop a Lorentzian pseudo-mode approximation to treat the non-Markovian evolution of an individual dipole qubit embedded in the quantized near field of a plasmonic nanoparticle. The approach admits analytical solutions that are used to follow the frequency-time dynamics of the single photon creation process and to generalize the strong coupling criterion. We show that the dynamics and spectral response of the near field can be tuned by controlling the width of the spectral distribution of the coupled qubit-photon state, relative to the intrinsic width of the plasmonic cavity spectrum, and suggest criteria for completely suppressing the Rabi splitting in strong coupling via single photon initialization. 

We analyze the photon creation process in strong coupling over femtosecond timescales. We find that the characteristic timescale for establishing a Rabi splitting in the coupled field spectrum is intimately related to the shape of the nanocavity spectrum. Before the Rabi splitting emerges, a dipole-emitted single photon is shown to interfere with itself in frequency and time, due to wavepacket evolution that is characteristic of coherent quantum systems that can be phase controlled \cite{Kohler1995,Shapiro2003,Ohmori2009}. Our results suggest new methods for manipulating the quantum state of single photons using currently available optical techniques.

The rest of the article is organized as follows: In Sec.~\ref{sec:macroscopic qed}, we review the macroscopic QED framework used to treat the non-Markovian system dynamics. In Sec. \ref{sec:LorentzianFit}, we describe the Lorentzian pseudo-mode approximation for the light-matter memory kernel. In Sec. \ref{sec:dipole on sphere}, we validate the approximation for a dipole near a metal nanoparticle. In Sec. \ref{sec:excitedphoton}, we study the dependence of the qubit evolution on the frequency content of the initial photon wavepacket, and in Sec. \ref{sec:interference} we study the phenomenon of single-photon interference. Conclusions are presented in Sec.~\ref{conclusion}.

\section{Macroscopic QED Framework}
\label{sec:macroscopic qed}

Light-matter interaction is treated using macroscopic QED theory \cite{Buhmann2009}. The dipole emitter is described as a charge distribution and the cavity is defined by the confined near field of an arbitrary material structure with local linear permittivity $\epsilon(\omega)$ and permeability $\mu(\omega)$. The total Hamiltonian derived from first principles can be written as \cite{Feist2021, Wang2019, Medina2021, Schafer2024}
 \begin{equation}\label{eq:total H}
\hat{\mathcal{H}}=\hat{\mathcal{H}}_{\text{dip}}+\hat{\mathcal{H}}_\text{f}+\hat{\mathcal{H}}_{\text{int}} + \hat{\mathcal{H}}_{\text{D}}(t),
\end{equation}
with the qubit described by $\hat{\mathcal{H}}_{\text{dip}}=\hbar\omega_{g} \lvert g\rangle\langle g\rvert+\hbar\omega_{e}\lvert e\rangle\langle e\rvert$, where $\lvert g\rangle$ and $\lvert e\rangle$ are the ground and excited dipole states, with frequencies $\omega_g$ and $\omega_e$, respectively.  The quantized near field is described by %
 \begin{equation}\label{eq:field H}
\hat{\mathcal{H}}_\text{f}=\int_0^\infty d\omega\,\hbar\omega\;\hat{a}^\dagger(\omega)\hat{a}(\omega), 
 \end{equation}
with field operators satisfying bosonic commutation relations $[\hat{a}(\omega),\hat{a}^\dagger(\omega')]=\delta(\omega-\omega')$ and $[\hat{a}(\omega),\hat{a}(\omega')]=0$. These are defined from the fundamental field modes $\hat{\vb{f}}_\lambda(\vb{r}_i,\omega)$ in macroscopic QED \cite{Buhmann2009}, through an orthonormalization procedure introduced in Ref. \cite{Feist2021}, which is reviewed in  Appendix \ref{app:orthonormalization} for completeness.

The light-matter interaction Hamiltonian is given by
 \begin{equation}\label{eq:light-matter H}
\hat{\mathcal{H}}_{\text{int}}=\int_0^\infty d\omega \hspace{2pt} g(\omega) \hat{d}^{(+)} \hat{a}(\omega)+\text{H.c},
 \end{equation}
 with coupling function  (units of \mbox{J C$^{-1}$m$^{-1}$Hz$^{-{1}/{2}}$}) given by
 \begin{equation}\label{eq:g coupling}
 g(\omega)=\sqrt{\frac{\hbar \omega^2}{\pi \epsilon_0 c^2}\vb{e}_\alpha\cdot \Im[\stackrel{\leftrightarrow}{\vb{G}}(\vb{r}_0,\vb{r}_0,\omega)]\cdot \vb{e}_\alpha},
 \end{equation}
where $\stackrel{\leftrightarrow}{\vb{G}}(\vb{r}_0\vb{r}_0,\omega)$ is the dyadic Green's tensor \cite{Chew1999} from the solution of Maxwell's equations for a specific nanocavity geometry and dielectric function $\epsilon(\omega)$, with a dipole source located at $\mathbf{r}_0$ and oriented along $\mathbf{e}_\alpha$, with $\alpha =\{x,y,z\}$. The electric dipole operator is given by $\hat d= \hat d^{(+)}+\hat d^{(-)}$ with $\hat{d}^{(+)}=d_{eg} \lvert e\rangle\langle g\rvert$ and $\hat{d}^{(-)}=[\hat{d}^{(+)}]^\dagger$. $d_{eg}=\langle e\rvert \hat d\rvert g\rangle$ is the transition dipole moment. $\epsilon_0$ is the permittivity of free space and $c$ the speed of light.  

We include in Eq. (\ref{eq:total H}) a phenomenological driving term of the form
 \begin{equation}\label{eq:driving H}
\hat{\mathcal{H}}_{\text{D}}(t)=\int_0^\infty d\omega \hspace{2pt}\hbar D(\omega)\hat{a}(\omega)\mathrm{e}^{i\omega t}+\text{H.c}
\end{equation}
where $D(\omega)$ is the driving strength parameter (units of Hz$^{1/2}$). From quantum optics theory \cite{Dorner2002}, $D(\omega)$ can be shown to scale with the far-field amplitude of a driving laser $E_D(t)$. This term injects photons into the nanocavity field, without directly driving the dipole.

To solve the the time-dependent Schrödinger equation  $i\hbar\partial_t |\psi(t)\rangle=\hat{\mathcal{H}}\lvert\psi(t)\rangle$ with the total Hamiltonian from Eq. (\ref{eq:total H}), we use the following Wigner-Weisskopf ansatz for the dipole-photon state $\lvert \psi(t)\rangle$
\begin{eqnarray}\label{eq:wf ansatz}
\lvert\psi(t)\rangle&=& C_{g0}(t)\lvert g\rangle\lvert\{0\}\rangle+C_{e0}(t)\mathrm{e}^{-i\omega_{e} t}\lvert e\rangle\lvert\{0\}\rangle\nonumber\\
&&+\int_0^\infty d\omega  C_{g1}(\omega,t)\mathrm{e}^{-i\omega t}\lvert g\rangle\lvert\{\vb{1}(\omega)\}\rangle
\end{eqnarray}
where $\lvert\{0\}\rangle$ is the photonic vacuum and $\lvert\{\vb{1}(\omega)\}\rangle=\hat a^\dagger(\omega)\ket{\{0\}}$ is the one photon eigenstate of $\hat{\mathcal{H}}_\text{f}$. The amplitude $C_{g1}(\omega,t)\equiv \langle 0|\hat a(\omega)|\psi(t)\rangle$ is the single-photon wavefunction \cite{Sipe1995,Fedorov2005}. 

Inserting Eq. (\ref{eq:wf ansatz}) in the Schrödinger equation gives the following integro-differential equation (IDE) for the dipole excitation amplitude
\begin{equation}\label{eq:Ce0}
\Dot{C}^{(e,0)}(t)=-\int_0^{t} dt^{\prime}\mathcal{K}(t-t')C_{e0}(t')+S(t),
\end{equation}
with the evolution governed by the delay kernel
\begin{equation}\label{eq:delay kernel}
\mathcal{K}(t-t')= \int_0^\infty d\omega\, \mathcal{K}(\omega) \mathrm{e}^{-i(\omega-\omega_{e})(t-t')}.
\end{equation}
The kernel spectrum $\mathcal{K}(\omega)=d_{eg}^2\lvert g(\omega)\rvert^2 /\hbar^2$ (units of Hz) is sometimes referred to as spectral density in the literature and denoted as $J(\omega)$ or $\kappa(\omega)$. For example $\mathcal{K}(\omega)\equiv(\sqrt{\epsilon_b}/8) J_\mu(\omega)$ in Ref. \cite{CuarteroGonzalez2020} and $\mathcal{K}(\omega)\equiv\kappa(\omega)$  in Ref. \cite{Hakami}.

The source term in Eq. (\ref{eq:Ce0}) is given by
\begin{eqnarray}\label{eq:source term}
\mathcal{S}(t)&=&-\frac{1}{\hbar}\int_0^\infty d\omega \sqrt{\mathcal{K}(\omega)} \left[ C_{g1}(\omega,0)\right.\nonumber \\
&&\left.-t   D(\omega) C_{g0}(0)\right]\mathrm{e}^{-i(\omega-\omega_e) t},
\end{eqnarray}
and is determined by the kernel spectrum, the initial state and the spectrum of the driving field. To complete the solution, Eq. (\ref{eq:Ce0}) must be supplemented by the single photon amplitude  
\begin{eqnarray}\label{eq:Cg1}
C_{g1}(\omega,t)&=&C_{g1}(\omega,0)-i t D(\omega)C_{g0}(0)\\
&&-i\int_0^{t} dt^{\prime}\;\sqrt{\mathcal{K}(\omega)}C_{e0}(t')\mathrm{e}^{i(\omega-\omega_{e})t'}\nonumber
\end{eqnarray}
and the dipole ground state amplitude
\begin{equation}\label{eq:Cg0}
C_{g0}(t)=C_{g0}(0)-i\int_0^t dt' \int_0^\infty d\omega\; D(\omega)C_{g1}(\omega,t').
\end{equation}

The derivation of  Eqs.~(\ref{eq:Ce0}), (\ref{eq:Cg1}) and (\ref{eq:Cg0}) is given in Appendix \ref{app:eom}. To preserve the norm of the single-excitation ansatz, to lowest order we only keep linear terms in $D(\omega)$.

\section{Lorentzian Kernel Approximation}
\label{sec:LorentzianFit}

In general, the IDE in Eq. (\ref{eq:Ce0}) can be solved numerically using standard integration methods \cite{LEATHERS2005}, once the delay kernel  $\mathcal{K(\tau)}$ is known. For specific types of kernels, analytical solutions can be obtained using Laplace transform techniques \cite{Polyanin2008handbook,BERMAN2010}. Consider an exponential memory kernel of the form $\mathcal{K}(\tau)=A\,{\rm exp}[-\tilde B\tau]$, where $A$ and $\tilde B$ are complex-valued constants. In the Laplace domain, Eq. (\ref{eq:Ce0}) becomes $C_{e0}(s)= \left[S(s) + C_{e0}(0)\right]G(s)$, where $S(s)$ is the Laplace transform of the source $S(t)$,  $C_{e0}(0)$ is the initial excited state amplitude, and $G(s)= (s+\tilde B)/(s^2+\tilde B s+A)$. For an initially excited qubit without sources, inverting back to the time domain gives 
\begin{equation}\label{eq:Ce0 analytical}
C_{e0}(t) =  {\rm e}^{-\tilde Bt/2}\left[\cos(bt)+\frac{\tilde B}{2b}\sin(bt)\right],
\end{equation}
with $2b=\sqrt{4A-\tilde B^2}$. The long-time stability of this solution imposes constraints on $A$ and $B$. Qualitatively, for a kernel with large enough amplitude and small  memory decay rate ($|A|\gg |B|^2/4$), the parameter $b$ is pure imaginary and Eq. (\ref{eq:Ce0 analytical}) describes damped Rabi oscillations, signature of strong coupling, with population lifetime $T_1=1/{\rm Re}[\tilde B]$ and Rabi frequency $\Omega_R=2{\rm Re}[b]$. If the kernel amplitude or  memory time are very small ($|A|\ll |B|^2/4$), then $C_{e0}$ undergoes simple exponential decay from its initial amplitude, signature of weak coupling. This intuitive picture is preserved when a source $S(t)$ is present. 

The generalization of Eq. (\ref{eq:Ce0 analytical}) is worked out by writing the memory kernel as the parametric model 
\begin{equation}\label{eq:memory multimode}
\mathcal{K}(\tau)=\sum_{j=1}^n A_j\,\mathrm{e}^{\left(-B_j-i[\Omega_j-\omega_e]\right)\tau}, 
\end{equation}
where $A_j>0$ is a real amplitude, $B_j>0$ a decay rate, and $\Omega_j$ is the frequency relative to $\omega_e$ of the $j$-th term. $n$ is the number of terms in the expansion. The spectrum of this model kernel is a sum of Lorentzians of the form
\begin{equation}\label{eq:kernel lorentzian}
\mathcal{K}(\omega)=\sum_{j=1}^n \frac{A_j}{\pi} \frac{B_j}{(\omega-\Omega_j)^2+B_j^2}, 
\end{equation}
where each Lorentzian is centered at $\Omega_j$ with bandwidth $2B$ (FWHM) and area $A_j$. This expansion is used to approximate the true kernel spectrum $\mathcal{K}(\omega)$ through a fitting procedure with $(n,A_j,B_j,\Omega_j)$ as fit parameters. The Lorentzian sum approximation for the kernel does not imply that the underlying photonic modes are non-interacting, as it occurs for uncoupled normal mode solutions \cite{Medina2021}. Instead, they can be interpreted as effective pseudo-modes of the photonic nanostructure that interact with the dipole qubit. 

If we also assume that the driving amplitude function $D(\omega)$ and the initial photon amplitude $C_{g1}(\omega,0)$ have peaked structure in the frequency domain, then the products $\sqrt{K(\omega)}D(\omega)$ and $\sqrt{\mathcal{K}(\omega)}  C_{g1}(\omega,0)$, which determine the source term $S(t)$ in Eq. (\ref{eq:source term}), can be fitted with other Lorentzian expansions analogous to Eq. (\ref{eq:kernel lorentzian}), using additional fit parameters. 

In this Lorentzian approximation, the solution of Eq (\ref{eq:Ce0}) can be written analytically as $ C_{e0}(t)=\sum_{j=1}^n \mathcal{X}_j\mathrm{exp}[-\mathcal{Y}_j t]$, where $\mathcal{X}_j$ and $\mathcal{Y}_j$ are given by numerical complex roots of high-order polynomials in the Laplace variable $s$. The explicit form of the polynomials is given in Appendix \ref{app:Laplace IDE}. The complex amplitudes and exponents ($\mathcal{X}_j$,$\mathcal{Y}_j$) are used to characterize the non-Markovian evolution of the dipole-nanocavity system.

\begin{figure}[b]
\includegraphics[width=0.41\textwidth]{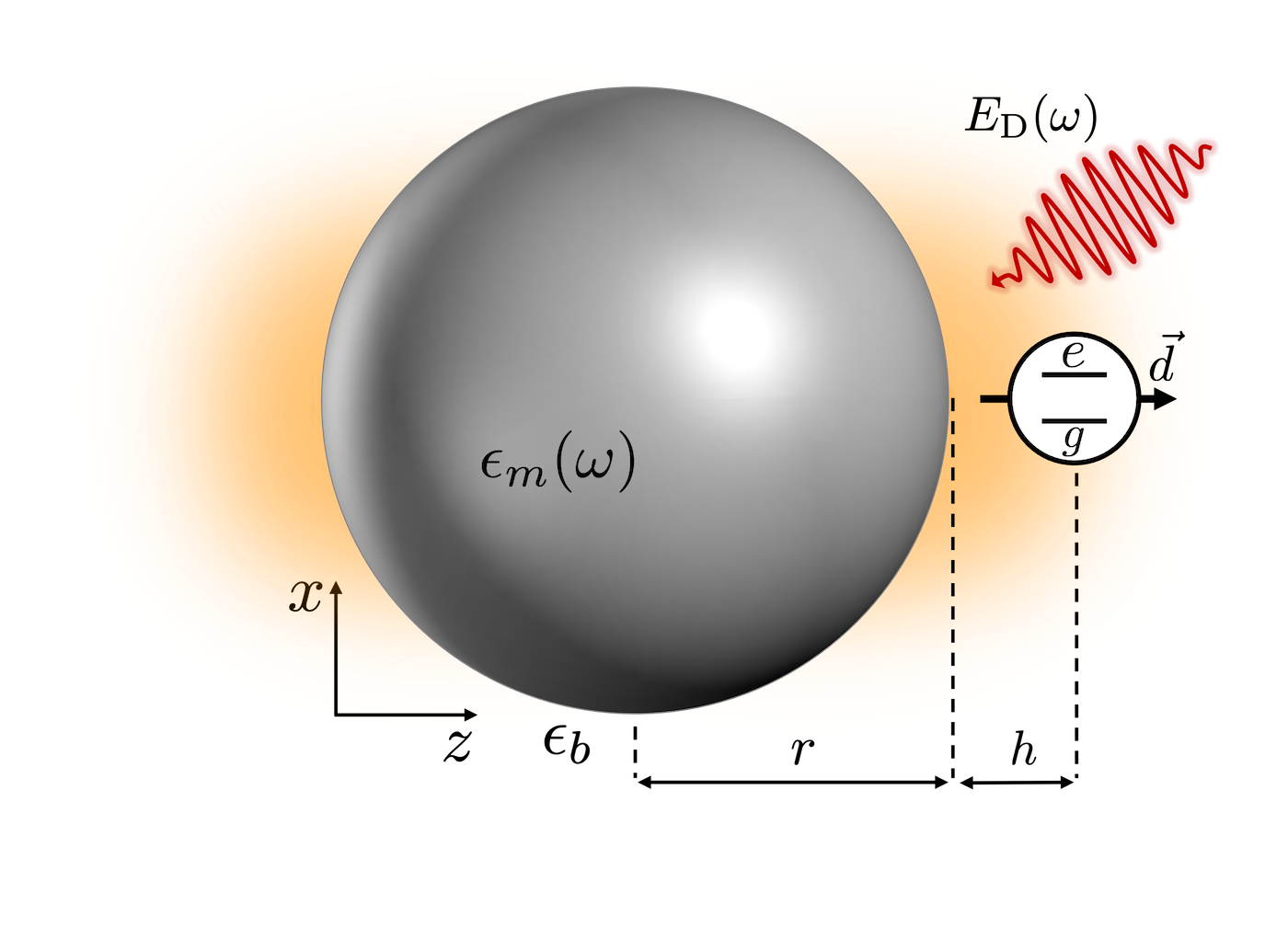}
\caption{Nanocavity scheme of a single two-level quantum emitter with transition dipole moment $\mathbf{d}$ oriented in the $z$-direction, at a distance $h$ from the surface of a metal nanosphere with dielectric function $\epsilon_m(\omega)$. The dipole is in the near field of the metal-dielectric interface. The system can be driven by a laser field $E_D(\omega)$.  $\epsilon_b$ is the background dielectric constant.}
\label{fig:scheme}
\end{figure}

\section{Non-Markovian Qubit-Nanocavity Dynamics}
\label{sec:dipole on sphere}

Figure \ref{fig:scheme} illustrates the dipole-nanocavity system studied here. An individual two-level dipole emitter with a large transition dipole moment $d_{eg}$ is placed at a distance $h$ from the surface of a silver nanosphere with radius $r$ and Drude permittivity $\epsilon_m(\omega)=\epsilon_\infty-\omega_p^2/(\omega^2+i\omega\Gamma)$. The nanocavity field corresponds to the confined optical near field at the metal-dielectric interface. We use $d_{eg}=24$ D (as in quantum dots \cite{Savasta2010}) and Drude parameters  $\omega_p=7.9$ eV,  $\Gamma=51$ meV, and  $\epsilon_\infty=6$, taken from \cite{Vanvlack2012}. The background medium is non-magnetic with $\varepsilon_b=1.0$. Other choices of dielectric background would introduce shifts to the kernel spectrum \cite{Zhao2008}, but these changes do not qualitatively affect the qubit-field interaction dynamics.

We use Mie theory \cite{Mie1908} to numerically construct the dyadic Green tensor of the sphere-dipole system from analytical   scattering amplitudes \cite{Yee1994}. Without losing generality, the dipole qubit is oriented along the $z$-direction at distances $h=2$ nm and $h=10$ nm from the surface of a silver sphere of radius $r = 20$ nm. From the tensor element $G_{zz}$ we compute the exact kernel spectrum $\mathcal{K}(\omega)\propto |g(\omega)|^2$, using Eq. (\ref{eq:g coupling}) for the light-matter coupling function $g(\omega)$.  

\begin{figure}[t]
\includegraphics[width=0.45\textwidth]{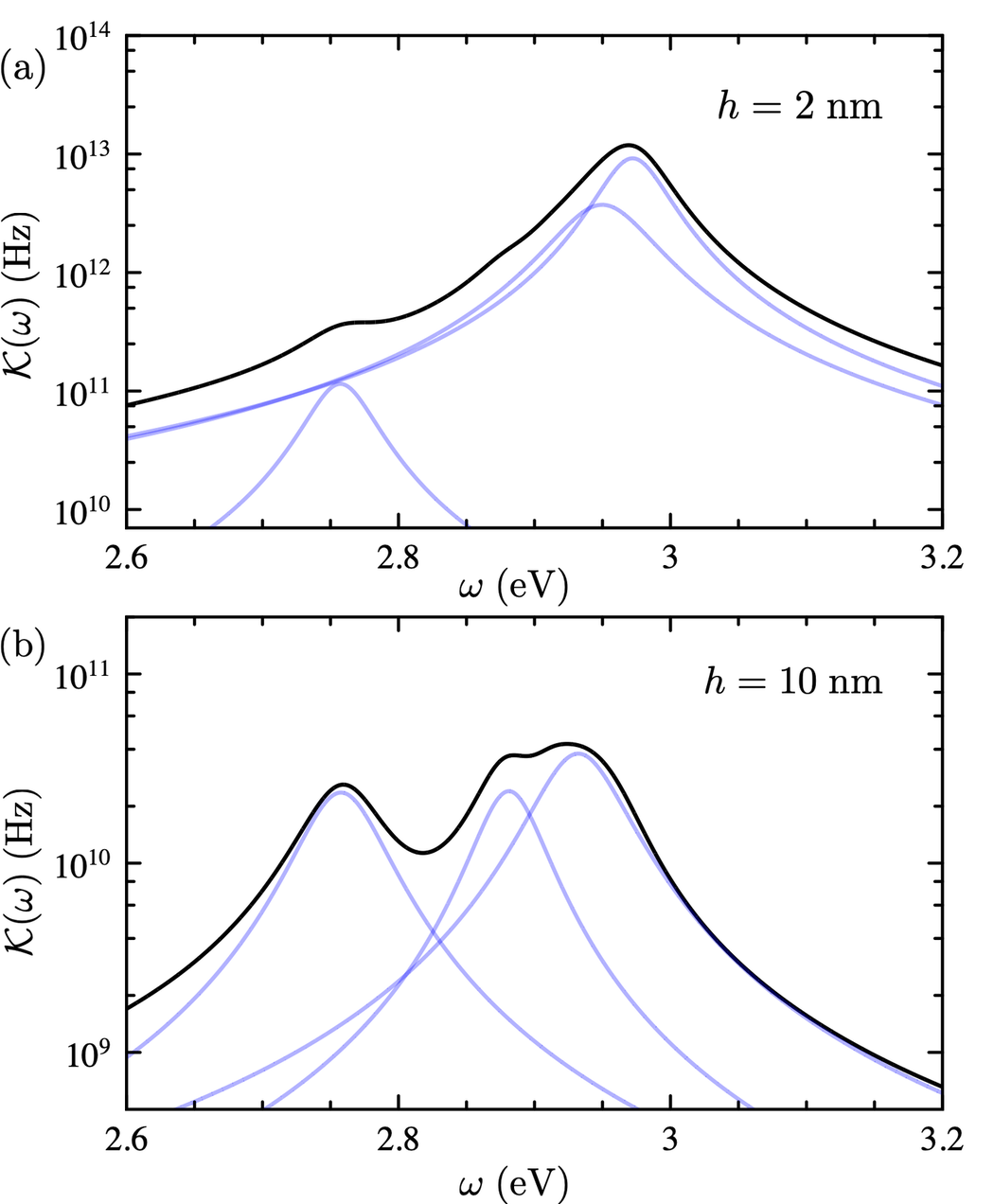}
\caption{(a) Exact kernel spectrum $\mathcal{K}(\omega)$ (thick black line) for a qubit dipole separated by $h=2$ nm from the surface of the silver nanosphere ($r=20$ nm). Individual contributions to the Lorentzian kernel approximation (light blue) are also shown. (b) Same as panel (a) for $h=10$ nm.}
\label{fig:kernels}
\end{figure}

Figure~\ref{fig:kernels} shows the kernel spectrum $\mathcal{K}(\omega)$ (thick solid line) for $h= 2$ nm and $h=10$ nm, together with the best fitted Lorentzians (thin solid lines) that reproduce the spectral features with $n=3$ pseudo-modes. In principle the number of Lorentzians $n$ can arbitrary, but we use $n=3$ following previous work \cite{Vanvlack2012,CuarteroGonzalez2020} showing that in order of increasing frequency this system has a low-frequency dipole mode ($\sim 2.8$ eV in Fig. \ref{fig:kernels}, bottom), a quadrupole mode ($\sim2.87$ eV) and several higher-order multipole modes that tend to cluster around at higher frequencies ($\sim 2.97$ eV). By decreasing the dipole-sphere distance $h$, at fixed radius, the multipole modes tend to dominate the kernel spectrum. Increasing $h$ reduces the overall magnitude of $\mathcal{K}$ and decreases the relative contribution of the multipole features relative the dipole peak.

\begin{figure}[t]
\includegraphics[width=\columnwidth]{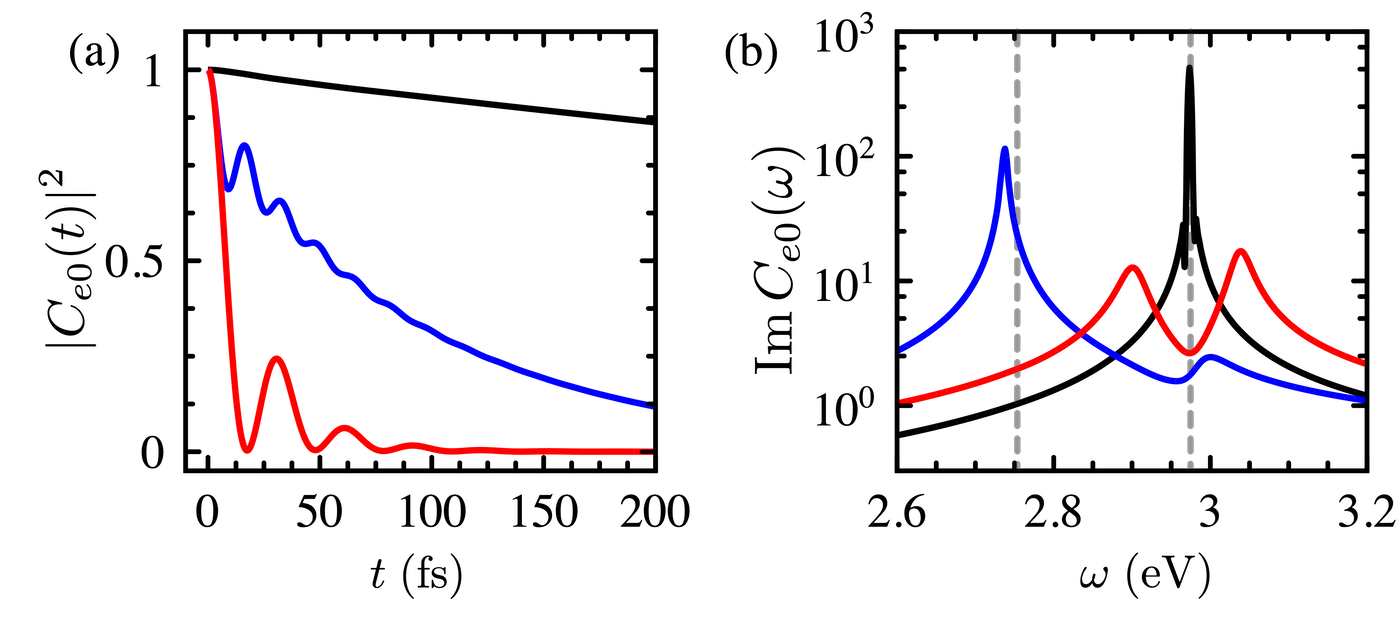}
\caption{(a) Excited population dynamics for qubit dipole separated by $h=2$ nm (red and blue lines) and $h=10$ nm (black line) from the silver sphere surface. For $h=2$ nm, the qubit is resonant ($\omega_e=2.97$ eV, red curve) or detuned ($\omega_e=2.75$ eV, blue curve) from the maximum of the kernel spectrum; (b) Fourier spectra of the imaginary part of the excited-state coherence, corresponding to the population traces in panel (a). Vertical dashed lines mark the qubit transition frequencies.}
\label{fig:excited dipole}
\end{figure}

To validate the proposed Lorentzian scheme for solving Eq. (\ref{eq:Ce0}), we show in Fig. ~\ref{fig:excited dipole}a the excited state population dynamics of an initially excited qubit [$C_{e0}(0)=1$] in the absence of driving [$S(t)=0$], comparing results for two qubit frequencies $\omega_e$ and two separation distances $h$. Figure \ref{fig:excited dipole}b shows the corresponding Fourier spectra of the population time traces. The spectra quantitatively agrees with the results in Ref. ~\cite{Vanvlack2012}. The field is weakly coupled to the excited dipole when $\omega_e$ is detuned from the multipole feature at small distances ($h=2$ nm), and spontaneous emission occurs with a Purcell factor $\gamma/\gamma_0\sim 10^3$. $\gamma_0$ is the free space qubit emission rate. Tuning the qubit to the multipole spectral feature leads to damped ultrafast Rabi oscillations and Rabi splitting in the Fourier domain. Figure \ref{fig:excited dipole} also shows that for $h=10$ nm, only Purcell-enhanced emission is expected ($\gamma/\gamma_0\sim 700$).

\begin{figure*}[t]
\includegraphics[width=0.9\textwidth]{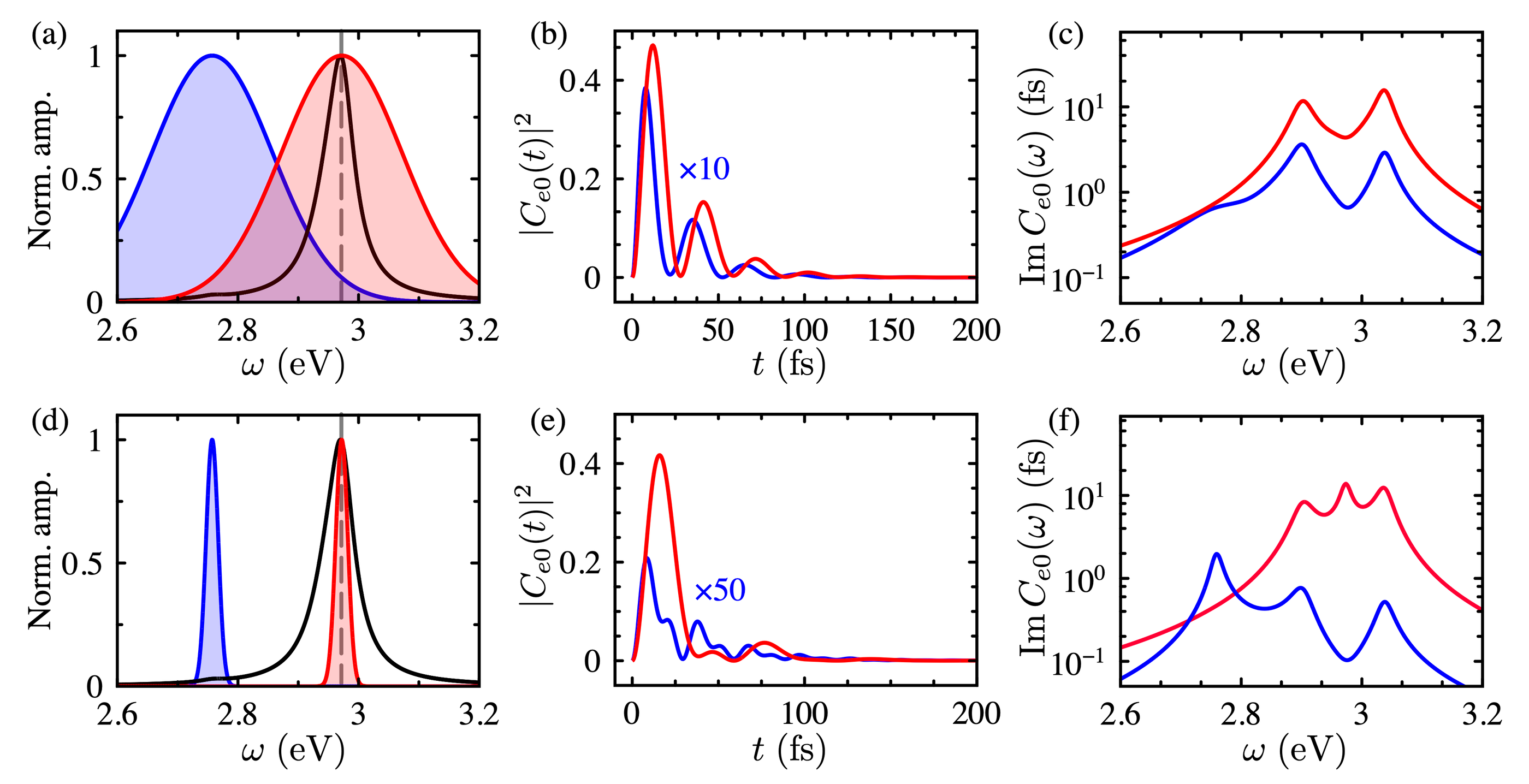}
\caption{(a) Initial single photon Gaussian amplitudes $C_{g1}(\omega)$ ($\sigma=0.1$ eV) on resonance (red line) and far detuned (blue line) from the maximum of the kernel spectrum (black line). The vertical dashed line marks the qubit frequency ($\omega_e=2.97$ eV);  (b) Excited population dynamics associated with the initial photon amplitudes in panel (a). The detuned dipole population (blue line) is scaled by a factor of 10, for clarity; (c) Fourier spectra of the imaginary part of the excited dipole coherence corresponding to the curves in panel (b); Bottom panels (d), (e), and (f) describe the same as the top panels, for narrower initial photon distributions with $\sigma=0.01$ eV. The dipole-surface distance is $h=2$ nm.} 
\label{fig:initial photon}
\end{figure*}

Additional comparisons with numerically exact qubit population results for a nanoparticle-on-mirror (NoM) geometry from Ref. \cite{CuarteroGonzalez2020} are carried out in Appendix \ref{app:comparison}. For this nanocavity system, the Lorentzian kernel approximation underestimates the Rabi period by about $2 \,{\rm fs}$ in strong coupling, and overestimates the spontaneous emission lifetime by about $10-60\, {\rm fs}$ in weak coupling. NoM geometries confine the near field to picometer scale mode volumes and are extensively used in strong coupling experiments using few-Debye molecular dipoles \cite{Chikkaraddy2016,Wang2019,Neuman2020,Esteban2022,Roelli2024}.

\section{Controlled Qubit Dynamics via single-photon shaping}
\label{sec:excitedphoton}

\subsection{Initially excited photon}
Having validated our methodology using initially excited dipoles, we now discuss a scenario mostly ignored in the literature, where the qubit is initially in the ground state and the photon field is excited. This is qualitatively different from the excited qubit picture because the near-field photon is distributed over a broad range of frequencies according to the probability distribution $|C_{g1}(\omega)|^2$. We ensure strong coupling by placing a resonant qubit ($\omega_e=2.97$ eV) close to the metal surface ($h = 2$ nm).

For  a single photon wavepacket produced in the distant past with Gaussian amplitude, the initial photon state is defined by
\begin{equation}\label{eq:gaussian}
C_{g1}(\omega,0) = A_0\, {\rm e}^{-\frac{(\omega - \omega_s)^2}{2\sigma^2}}  
\end{equation}
with amplitude $A_0$, width parameter $\sigma$, and central frequency $\omega_s$. The qubit is initially in the ground state.  Normalization of the wave function gives $A_0 = 1/\sqrt{\sigma \sqrt{\pi}}$.

Figure \ref{fig:initial photon}(a) shows two broad single photon distributions with $\sigma=0.1$ eV. One distribution (red curve) is centered at the peak of the kernel spectrum $\mathcal{K}(\omega)$ at 2.97 eV (dashed vertical line), overlapping completely with the multipole peak.  The other distribution (blue curve) is red detuned from the peak of the kernel by about $0.2$ eV, but the high frequency tail strongly overlaps with the multipole peak. Panels \ref{fig:initial photon}(b) and \ref{fig:initial photon}(c) show the corresponding qubit population evolution and dipole coherence spectrum, respectively. For the resonant distribution (red curve), the excited dipole undergoes Rabi oscillations of the population and the coherence spectrum exhibits Rabi splitting $\Omega_R\approx 0.134$ eV. Rabi oscillations are preserved when the photon distribution is detuned from the kernel peak, although with smaller amplitude (blue curve), since only the tail of the photon distribution overlaps with the main kernel feature. In this case, the coherence spectrum shows a higher relative contribution at the center photon frequency $\omega_s\approx 2.75$ eV, corresponding to resonance fluorescence in weak coupling. Both elastic (resonance fluorescence) and inelastic (Rabi splitting) qubit responses coexist. 

To assess the impact of the distribution bandwidth $\sigma$ on the qubit dynamics, we repeat the analysis for distributions that are centered at the same frequencies ($2.97$ eV resonant, $2.75$ eV red-detuned), but are ten times narrower, i.e., $\sigma = 0.01$ eV.  Figure \ref{fig:initial photon}(d) shows the resonant (red line) and far-detuned (blue line) Gaussian distributions superposed on the kernel spectrum. In this case, the resonant distribution only partially overlaps with the kernel spectrum and the red-detuned distribution has an insignificant tail overlap with the kernel peak. Figure \ref{fig:initial photon}(e) shows that in both cases the qubit is less efficiently excited and Rabi oscillations are less regular. The off-resonant population trace is significantly weaker. The qubit coherence spectrum in Fig. \ref{fig:initial photon}(f) shows that the narrow elastic response dominates over the strong coupling Rabi doublet, which is greatly suppressed in peak strength. The Rabi splitting $\Omega_R\approx 0.130$  eV is roughly the same as in Fig. \ref{fig:initial photon}(c), confirming the intuition from Eq. (\ref{eq:Ce0 analytical}) that it is largely determined by the kernel.

\begin{figure}[t]
\includegraphics[width=0.45\textwidth]{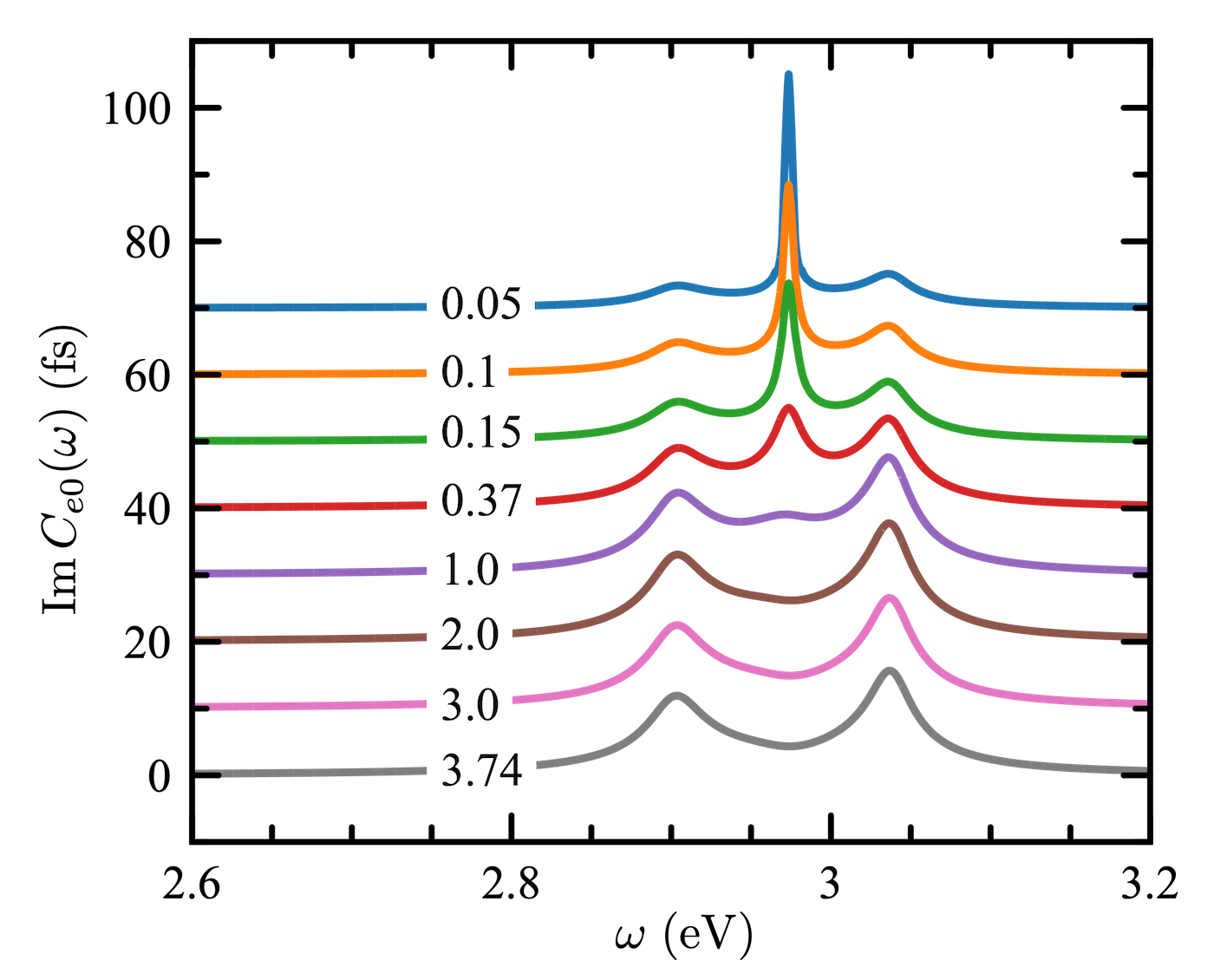}
\caption{Fourier spectra of the dipole coherence for Gaussian single photon distributions of different bandwidths, all centered at the qubit frequency ($\omega_s=\omega_e=2.97$ eV). Curves are labelled by the bandwidth (FWHM) ratio $\Gamma_G/\Gamma_K$. The dipole-surface distance is $h=2$ nm.}
\label{fig:width ratio}
\end{figure}

Figure~\ref{fig:width ratio} shows a more systematic analysis of the dipole coherence spectrum for different ratios between the full width at half maximum (FWHM) of the photon distribution $\Gamma_{\rm G} \approx 2.355\, \sigma $ and the FHWM of the kernel spectrum $\Gamma_{\rm K}= 63$ meV. The initial photon distribution is centered  at $\omega_s=\omega_e=2.97$ eV. For $\Gamma_{\rm G}/\Gamma_{\rm K}\ll 1$, the elastic scattering peak at the qubit frequency dominates over the Rabi shoulders, which is reminiscent of the Mollow triplet spectrum in atomic resonance fluorescence \cite{Mollow1969,Ng2022}. Beyond the crossover ratio $\Gamma_{\rm G}/\Gamma_{\rm K}\approx 1$,  the central elastic response is suppressed and the spectrum exhibits a well-defined doublet separated by $\Omega_R\approx 0.134$ eV. 

\subsection{Laser driving from the ground state}
We now consider the dipole response of the coupled system initially in the absolute ground state (no photons or qubit excitations), subject to an external laser field that injects energy into the near field of the nanoparticle through the driving term $\hat{\mathcal{H}}_{\rm D}$ in Eq. (\ref{eq:driving H}). The dipole moment of typical qubits are smaller than the dipole of metallic nanoparticles  \cite{effdipoleMNP}, so we neglect direct laser driving of the qubit. For consistency, we assume the driving amplitude has the Gaussian form $D(\omega)=D_0 {\rm exp}[-(\omega - \omega_s)^2/2\sigma^2]$, with center frequency $\omega_s$ and standard deviation $\sigma$. The amplitude $D_0$ is in principle a free parameter, but to remain within the single-excitation subspace, we restrict $D_0$ to values that preserve the state norm $|\langle \psi(t)|\psi(t)\rangle|^2\approx 1$ for times much longer than the Rabi period ($t\gg 1/\Omega_R$). We have numerically checked that for up to $D_0\sim 10^{-5} {\rm Hz}^{1/2}$, norm loss is smaller than 0.1\% up to $~300$ fs, for widths in the range $\sigma \sim 10-100$ eV.  

\begin{figure}[t]
\includegraphics[width=0.5\textwidth]{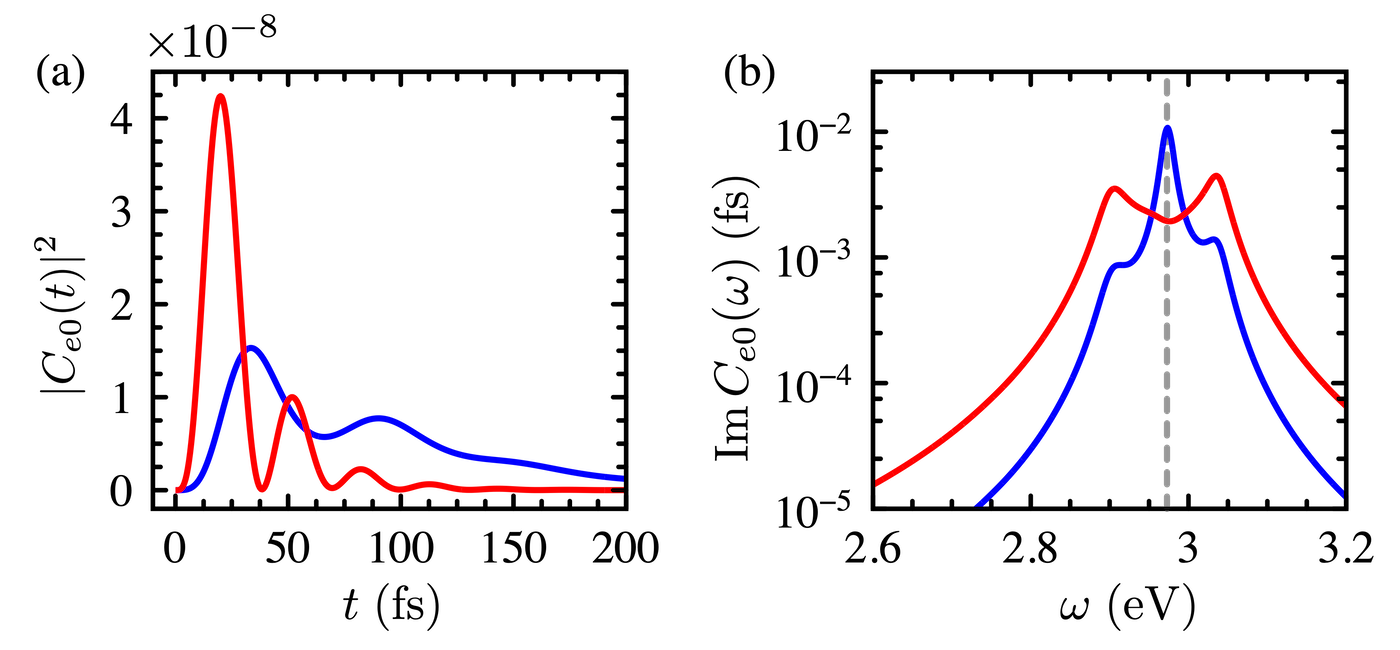}
\caption{(a) Excited population dynamics for a dipole strongly coupled to the nanocavity field ($h= 2$ nm), subject to a broadband ($\sigma = 0.1$ eV, red curve) and narrowband ($\sigma = 0.01$ eV, blue curve) laser field centered at the qubit frequency $\omega_e=2.97$ eV; (b) Fourier spectrum of the dipole coherence associated with the population traces in panel (a).  The system is initially in the absolute ground state and the driving strength is $D_0=811$ Hz$^{1/2}$.}
\label{fig:laser driving}
\end{figure}

\begin{figure*}[t]
\includegraphics[width=0.9\textwidth]{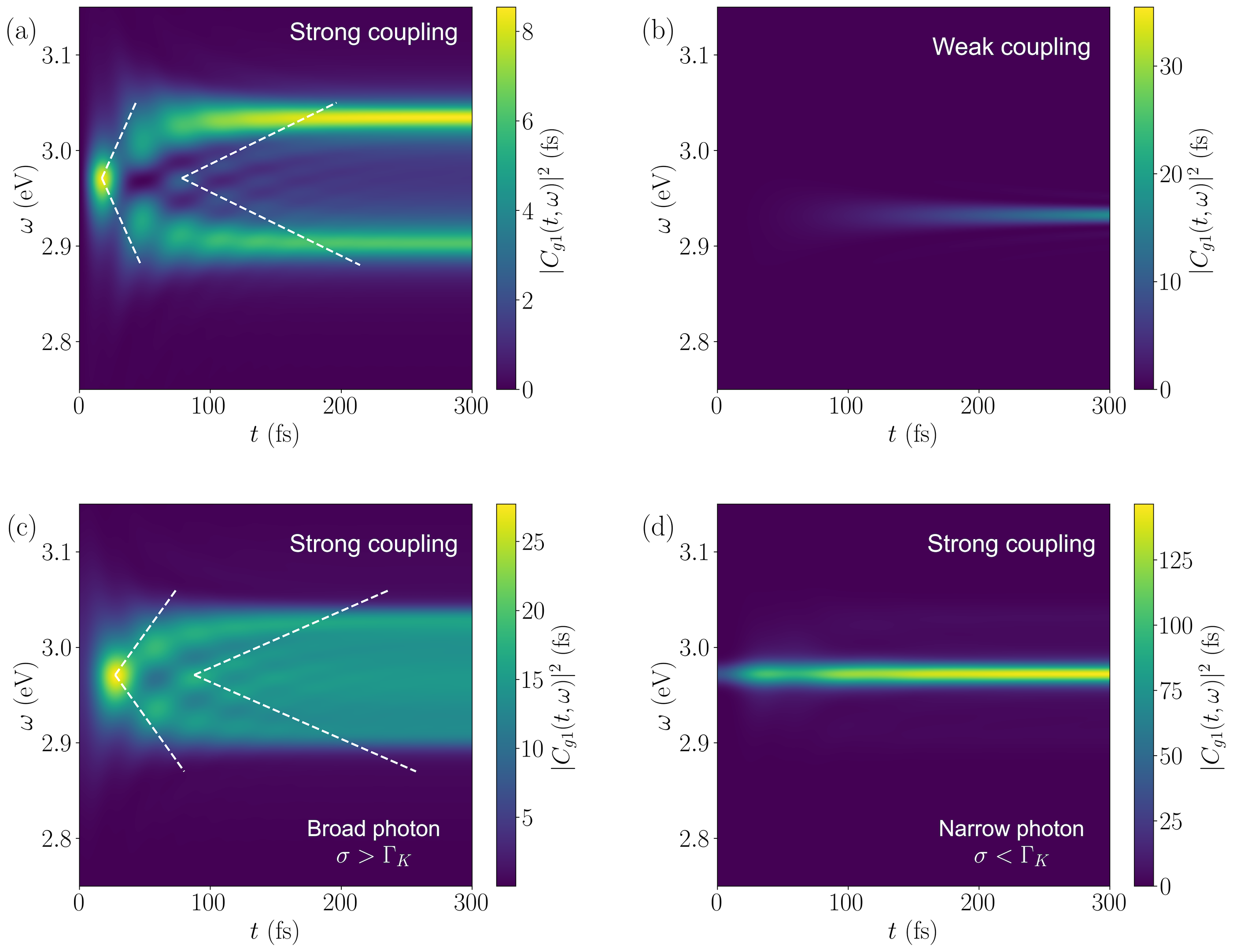}
\caption{(a) Single photon probability distribution in frequency and time, for an initially excited dipole with frequency $\omega_=2.97$ eV separated by $h=2$ nm from the sphere surface, which are strong coupling conditions. Analytical phase trajectories are marked with dashed lines; (b) Same as panel (a) for $h = 10$ nm, which gives weak coupling; (c) Single photon probability distribution for the same dipole-sphere system in panel (a), but with an initially excited photon having a broad Gaussian profile ($\sigma = 0.1$ eV). Analytical phase trajectories are also marked; (d) Same as panel (c) for a narrower initial photon amplitude ($\sigma = 0.01$ eV).}
\label{fig:Cg1 density}
\end{figure*}

Figure \ref{fig:laser driving}(a) shows the qubit population dynamics and Fig. \ref{fig:laser driving}(b) the corresponding dipole coherence spectrum, for narrow ($\sigma = 0.01$ eV) and broadband ($\sigma=0.1$ eV) driving amplitudes $D(\omega)$, with $D_0$ fixed. The results are qualitatively similar to having an initially excited photon (see Fig. \ref{fig:initial photon}), i.e., spectrally broad driving amplitudes lead to better resolved Rabi oscillations in the population dynamics and well-defined Rabi splittings in the spectral response. The width of the kernel spectrum $\Gamma_{\rm K}$ is the relevant parameter for comparing bandwidths. For narrow laser fields ($\sigma\ll \Gamma_{\rm K}$),  the elastic response dominates over the Rabi sidebands. Since $D_0$ remains small to preserve the state norm, ground state depletion is also perturbatively small. Extensions of the equations of motion beyond the single-excitation regime should be developed to model laser-induced ground state depletion.

\section{Ultrafast single-photon wavepacket interference}
\label{sec:interference}

In our Wigner-Weisskopf analysis of the coupled state, the electric field amplitude $E(\omega,t)\equiv \langle \hat E^{(-)}\rangle$ can be shown to depend on single-photon wavefunction $\langle 0|\hat a(\omega,t)|\psi(t)\rangle = C_{g1}(\omega,t)$, and the intensity $I(\omega,t)=\langle \hat E^{(+)}\hat E^{(-)}\rangle$ on the probability density $|C_{g1}(\omega,t)|^2$ (see derivation in Appendix \ref{app:eom}). Therefore, by analyzing the single photon wavefunction, we can understand the origin of strong coupling for qubits in optical nanocavities. 

Solving for $C_{g1}$ in Eq. (\ref{eq:Cg1}) using the analytical form in Eq. (\ref{eq:Ce0 analytical}) with $\Omega=\omega_e$, we obtain
\begin{equation}\label{eq:Cg1 analytical}
C_{g1}(\delta,t)=C_{g1}^\infty(\delta)\left(1-{\rm e}^{-\frac{B}{2}t+i\delta t}\left[\cos(bt)-p(\delta)\sin(bt)\right]\right),
\end{equation}
with detuning $\delta=\omega-\omega_e$, $p(\delta)\approx b/(B-i\delta)$ and the stationary amplitude
\begin{equation}
C_{g1}^\infty(\delta) = -i\sqrt{\mathcal{K(\delta)}}\,\frac{B-i\delta}{A-\delta^2-iB\delta}.
\end{equation}
The general expressions for $C_{g1}(\delta,t)$ and $p(\delta)$ for arbitrary qubit and kernel frequencies are given in Appendix \ref{app:Laplace IDE}.

For long times $t\gg 2/B$, the single photon intensity is determined by
\begin{equation}\label{eq:steady photon population}
|C_{g1}^\infty(\delta)|^2= \mathcal{K}(\delta)\frac{\delta^2+B^2}{A^2+(B^2-2A)\delta^2+\delta^4},
\end{equation}
showing Rabi splitting when $2A>B^2$. Otherwise, the spectrum has a single peak at the qubit frequency. Therefore, in terms of the total area and FWHM of the Lorentzian kernel $\mathcal{K}(\omega)$, a general strong coupling criterion for nanocavities can be formulated as
\begin{equation}
2{\int}d\omega \mathcal{K}(\omega)> (\Gamma_{K}/2)^2.
\end{equation}

 For intermediate times $1/b\gtrsim t\ll 1/B $, the single photon probability can be written as
 \begin{equation}\label{eq:Cg1 probability}
 |C_{g1}(\delta,t)|^2\approx|C_{g1}^\infty(\delta)|^2\left(1+|h(\delta,t)|^2-2{\rm Re}[h(\delta,t)]\right)
 \end{equation}
with $h(\delta,t)\equiv[\cos(bt)-p(\delta)\sin(bt)]{\rm e}^{i\delta t}$. The photon probability density has oscillation patterns in frequency-time space over timescales limited by the qubit relaxation time $T_1=1/B$. The last term in Eq. (\ref{eq:Cg1 probability}) gives  destructive and constructive interference, establishing trajectories of constant phase defined by 
\begin{equation}\label{eq:phase trajectory}
\frac{d}{dt}\left[(\delta(t) \pm b) t\right] = 0.
\end{equation}
Nodes or antinodes in the probability density at $(t_0,\delta_0)$ are sources of phase propagation in frequency-time, with trajectories following straight lines of slope $d\delta/dt= (\delta_0\pm b)/t_0$.  These interference patterns are the frequency-time analogues of spacetime quantum carpets known for multi-level systems \cite{Kaplan:1998,Berry2022}.

Figure \ref{fig:Cg1 density}(a) shows the single-photon distribution in frequency-time  associated with an initially excited qubit dipole that is resonant with the peak of the kernel spectrum for $h=2\,{\rm nm}$ above the silver surface. The density plot is obtained by numerically solving Eq. (\ref{eq:Ce0}) using the Lorentzian kernel approximation (fitting parameters in Table \ref{tab:fitparams}, Appendix \ref{app:Laplace IDE}). This scenario supports strong coupling (see Fig. \ref{fig:excited dipole}). The photon distribution has the features described in the analytical model: (\emph{i}) coherent wavepacket propagation up to about $T_1= 127 \,{\rm fs}$ ($B=32.5\,{\rm meV}$); (\emph{ii}) symmetric Rabi splitting $\Omega_R\approx 135 \,{\rm meV}$ centered at the qubit frequency for $t\gg T_1$. The shape of splitting is asymmetric (lower peak is weaker) due to the overall asymmetry of the kernel spectrum. The numerical splitting is larger than the splitting predicted from Eq. (\ref{eq:steady photon population}), $\Omega_R=\sqrt{4A-2B^2}\approx 99\, {\rm meV}$ ($A=0.00299\,{\rm eV}^2$), and larger than the splitting inferred from Eq. (\ref{eq:Ce0 analytical}), $\Omega_R=2b \approx 104 \, {\rm eV}$. The analytical frequency-time phase propagation velocities (dashed lines) predicted from Eq. (\ref{eq:phase trajectory}) also agree well with the numerical phase trajectories. Figure \ref{fig:Cg1 density}(b) shows that when the system is in weak coupling ($h=10$ nm), Purcell-accelerated decay of the qubit population does not sustain single photon wavepacket interference at short times. 

For comparison, Fig. \ref{fig:Cg1 density}(c) shows the evolution of the single photon distribution for the same strong coupling conditions as in Fig. \ref{fig:Cg1 density}(a), but assuming an initially excited photon with Gaussian amplitude having  $\sigma= 100\,{\rm meV}>\Gamma_K$. The qubit is initially in the ground state. The photon distribution also develops phase interference patterns at short times $t<T_1$ with the same phase propagation velocities as in Fig. \ref{fig:Cg1 density}(a), but the long-time Rabi splitting is less sharply defined. These results confirm that  single-photon interference is a feature of the kernel (area and bandwidth), not the initial state. 

Figure \ref{fig:Cg1 density}(d) shows that the single-photon quantum interference pattern can also be suppressed by tuning the bandwidth of the initial photon distribution. The same strong coupling conditions as in Fig. \ref{fig:Cg1 density}(c) are used, but the initial distribution is narrower ($\sigma = 10\,{\rm meV}<\Gamma_K$). This ability to modulate the frequency-time photon distribution by shaping the spectral content of the initial photon state suggests opportunities for exploring wavepacket quantum control techniques using femtosecond pulses \cite{Kohler1995,Kawashima1995,Shapiro2003,Ohmori2009}.

\begin{figure}[t]
\includegraphics[width=\columnwidth]{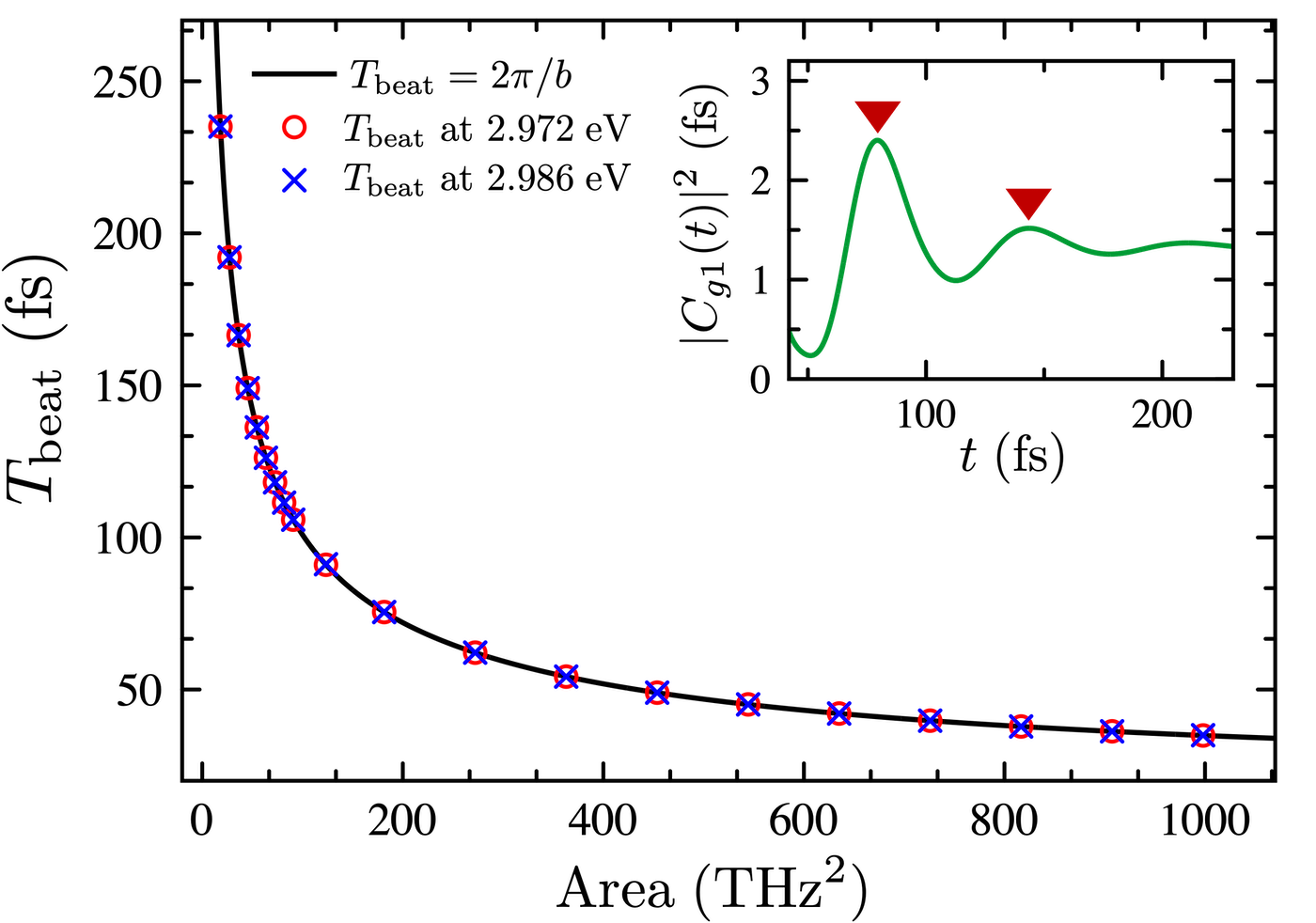}
\caption{Beating period $T_{\rm beat}$ of the single photon probability distribution as a function of the area of a model Lorentzian kernel with the same height as the true kernel $\mathcal{K}(\omega)$. The analytical scaling (solid line) is compared with numerical results from frequency cuts of the probability distribution at $2.972$ eV (circles) and $2.986$ eV (crosses). Inset: Evolution of the probability distribution at $2.972$ eV, with arrows marking the definition of  $T_{\rm beat}$. The qubit is initially in the excited state.}
\label{fig:beatings}
\end{figure}

Figure \ref{fig:beatings} shows the beating period $T_{\rm beat}$ of two time traces taken from the single photon distribution at $2.972$ eV and $2.986$ eV, as a function of the area of a single Lorentzian model kernel that coincides with the true kernel $\mathcal{K}(\omega)$ at the multipole peak frequency $\omega_{\rm max}= 2.97\, {\rm eV}$, but has tunable area and linewidth parameters $A$ and $B$, keeping $A/B=\pi\mathcal{K}(\omega_{\rm max})$.  The inset shows one of the time traces with the beating period marked. Phase beatings are determined by the interference term $h(\delta,t)$ in Eq. (\ref{eq:Cg1 probability}), which oscillates with frequency $b/2\pi$, giving $T_{\rm beat}\sim 1/b\sim 1/\sqrt{{\rm area}}$, where the area dependence comes from $b\sim \sqrt{A}$. For reference, the Lorentzian kernel with the same FWHM as the true kernel has area $\int \mathcal{K}(\omega)d\omega = 278.9 \,{\rm THz}^2$. Beating times $T_{\rm beat}\approx 100\, {\rm fs}$ should be detectable with available ultrafast heterodyne near-field probe techniques \cite{Park2019}.

\section{\label{conclusion}Conclusions and Outlook}

We developed a semi-analytical approximation to solve the non-Markovian dynamics of a two-level electronic dipole in an optical nanocavity, based on a macroscopic QED description of the broadband structure of the confined electromagnetic field. The approximation admits analytical solutions for the qubit and photon population and coherence that are valid at all timescales for arbitrary light-matter coupling strengths within the rotating-wave approximation. The approximation is accurate for nanocavity geometries whose local density of photonic states is highly structured in the frequency domain. Nanocavities with nanosphere and nanoparticle-on-mirror geometries are used here as test cases, but other nanocavity geometries and material compositions can be studied with the same methodology.

The coupled system evolution is unitary despite the broadband nature of the photonic spectrum. Depending on where the qubit frequency is located in the spectrum of the coupling kernel, Purcell-enhanced spontaneous emission in weak coupling or damped Rabi oscillations in strong coupling can be obtained. Based on the Lorentzian approximation, we formulate a general strong coupling criterion based of the form of the kernel spectrum at the qubit frequency. Non-radiative photonic relaxation is included in the theory through the complex dielectric function of the metallic nanocavity structure, which leads to photon absorption. However, electric dipole relaxation is ignored, limiting applicability of our results to describe dipole emitters with competing relaxation processes such quantum dots with electron-phonon scattering \cite{Vagov2002,PREZHDO2008} and organic dipole emitters that transfer charge \cite{Somoza2023,Panhans2023}. To model particular physical implementations, the macroscopic QED approach used here should be expanded with coherent or dissipative interaction terms that couple electrons with other material degrees of freedom \cite{Roy-Choudhury2015}. 

Our results demonstrate that commonly used treatments of light-matter interaction in optical nanocavities based on single-mode cavity QED theory \cite{Semenov2019,Litinskaya2019,Herrera2020,Herrera2022} are fundamentally limited in their ability to describe photonic observables over sub-picosecond timescales, which are experimentally accessible using ultrafast spectroscopy \cite{Zhao2025}. Measuring the predicted signatures of single photon interference in the frequency domain would open new perspectives towards controlling the photonic wavepacket dynamics with nanoscale resolution using quantum control techniques. 

\begin{acknowledgments}
The authors thank \'Alvaro Cuartero-Gonz\'alez and Antonio Fern\'andez-Dom\'inguez for sharing their data from Ref. \cite{CuarteroGonzalez2020}.  A.S.R. received support from USACH Excellence Graduate Scholarship. A.E.R.L. is supported by ANID-Fondecyt Iniciaci\'on No. 11250638. F.H. is supported by ANID-Fondecyt Regular Grant No. 1221420 and the Air Force Office of Scientific Research under Award No. FA9550-22-1-0245. All authors thank support from the ANID-Millennium Science Initiative Program No. ICN17\_012.

\end{acknowledgments}

\bibliography{singlephoton-macroQED-submit}

\newpage
\begin{widetext}

\appendix
\section{\label{app1}Orthonormalization of bosonic field operators}
\label{app:orthonormalization}

Consider a plasmonic environment with $N$ dipolar emitters in the near field. We write the cavity field hamiltonian in terms of the continuum of the fundamental bosonic field variables $\hat{\vb{f}}_\lambda(\vb{r},\omega)$ and $\hat{\vb{f}}^\dagger_\lambda(\vb{r},\omega)$ as,
\begin{equation}
 \hat{\mathcal{H}}_\text{f}=\sum_{\lambda=e,m}\int d^3\vb{r}'\int_0^\infty d\omega\hspace{3pt}\hbar\omega\hat{\vb{f}}^\dagger_\lambda(\vb{r}',\omega)\cdot\hat{\vb{f}}_\lambda(\vb{r}',\omega) \label{app1eq1}. 
\end{equation}
These bosonic operators satisfy the commutation relations $\left[\hat{\vb{f}}_\lambda(\vb{r},\omega),\hat{\vb{f}}_{\lambda'}(\vb{r}',\omega')\right]=\left[\hat{\vb{f}}^\dagger_\lambda(\vb{r},\omega),\hat{\vb{f}}^\dagger_{\lambda'}(\vb{r}',\omega')\right]=\vb{0}$ and $\left[\hat{\vb{f}}_\lambda(\vb{r},\omega),\hat{\vb{f}}^\dagger_{\lambda'}(\vb{r}',\omega')\right]=\delta_{\lambda\lambda'}\vb{\delta(r-r')}\delta(\omega,\omega')$. We model the interaction Hamiltonian in the electric dipole approximation as
\begin{equation}
  \hat{\mathcal{H}}_\text{int}=-\sum_{i=1}^N\hat{\vb{d}}_i(\vb{r})\cdot\hat{\vb{E}}(\vb{r})\label{app1eq2}
\end{equation}
where the dipole operator is given in terms of transition dipole matrix $ \hat{\vb{d}}_{i}(\vb{r})= \sum_\alpha \hat{d}_{i,\alpha}\cdot \vb{e}_\alpha$ in which $i$ and $\alpha$ in the subscripts represent the $i^{\text{th}}$ dipole and its orientation respectively. In macroscopic QED formalism, the electric field operator can be written in terms of dyadic Green's function,
\begin{equation}
  \hat{\vb{E}}(\vb{r})=\int_0^\infty d\omega \sum_{\lambda=e,m}\int d^3\vb{r}'\left[\stackrel{\leftrightarrow}{\vb{G}}_\lambda(\vb{r},\vb{r'},\omega)\cdot \hat{\vb{f}}_\lambda(\vb{r'},\omega)+\stackrel{\leftrightarrow}{\vb{G}}_\lambda^{*\intercal}(\vb{r},\vb{r'},\omega)\cdot \hat{\vb{f}}_\lambda^\dagger(\vb{r'},\omega)\right]\label{app1eq3}
\end{equation}
with expansion coefficients given by $ \tensor{\vb{G}}_e(\vb{r},\vb{r'},\omega)=ik_0^2\sqrt{\frac{\hbar}{\pi\epsilon_0}\Im[\epsilon(\vb{r'},\omega)]}\tensor{\vb{G}}(\vb{r},\vb{r'},\omega)$ for $\lambda=e$ and $ \tensor{\vb{G}}_m(\vb{r},\vb{r'},\omega)=ik_0\sqrt{\frac{\hbar}{\pi\epsilon_0}\frac{\Im[\mu(\vb{r'},\omega)]}{\abs{\mu(\vb{r'},\omega)}^2}}\left[\vb{\nabla'}\times\tensor{\vb{G}}_e(\vb{r},\vb{r'},\omega)\right]^\intercal$ for $\lambda=e$. The total Green's tensor, $\tensor{\vb{G}}(\vb{r},\vb{r'},\omega)$, is the solution of the Helmholtz wave equation. 
\par We can thus write the interaction hamiltonian in the rotating wave approximation as,
\begin{equation}
  \hat{\mathcal{H}}_\text{int}=\int_0^\infty d\omega \sum_i\sum_\alpha \sum_{\lambda=e,m}\int d^3r \left[\hat{d}_{i,\alpha}^{(+)}\vb{e}_\alpha\cdot\stackrel{\leftrightarrow}{\vb{G}}_\lambda(\vb{r},\vb{r}',\omega)\cdot\hat{\vb{f}}_\lambda(\vb{r}',\omega)+\hat{d}_{i,\alpha}^{(-)}\vb{e}_\alpha\cdot\stackrel{\leftrightarrow}{\vb{G}}_\lambda^{*\intercal}(\vb{r},\vb{r}',\omega)\cdot\hat{\vb{f}}_\lambda^\dagger(\vb{r}',\omega)\right]\label{app1eq4}
\end{equation}
We now look for a linear transformation of the bosonic modes $\hat{\vb{f}}_\lambda(\vb{r},\omega)$ at each $\omega$ to establish a minimal basis for the medium assisted EM field including the losses. These new emitter centered linearly independent modes encodes the complete information about the nanostructure cavity/antenna via the dyadic Green's tensor. Following \cite{Feist2021}, we introduce emitter-centered or bright EM modes $\hat{b}_i(\omega)$ associated with each emitter $i$ as
\begin{gather}
\hat{b}_{i,\alpha}\left(\omega\right)=\sum_\lambda \int d^3\vb{r}\hspace{2pt} \vb{u}_{i,\alpha,\lambda}(\vb{r},\omega)\cdot \hat{\vb{f}}_\lambda(\vb{r},\omega),\label{app1eq5}
\end{gather}
with
\begin{gather}
\vb{u}_{i,\alpha,\lambda}(\vb{r},\omega)=\frac{\vb{e}_\alpha\cdot \stackrel{\leftrightarrow}{\vb{G}}_\lambda(\vb{r}_i,\vb{r},\omega)}{\mathcal{E}_{i,\alpha}(\omega)}.\label{app1eq6}
\end{gather}
$\mathcal{E}_{i,\alpha}(\omega)$ is a normalization factor. The commutation relation $\hat{b}_{i,\alpha}\left(\omega\right)$ from the commutation relations of $\hat{\vb{f}}_\lambda(\vb{r},\omega)$ to give,
\begin{equation}
[\hat{b}_{i,\alpha}\left(\omega\right),\hat{b}_{j,\beta}^\dagger\left(\omega'\right)]=\sum_\lambda \int d^3\vb{r} \vb{u}^*_{i,\alpha,\lambda}(\vb{r},\omega)\cdot \vb{u}_{j,\beta,\lambda}(\vb{r},\omega')\left[\hat{\vb{f}}_\lambda(\vb{r},\omega),\hat{\vb{f}}^\dagger_{\lambda}(\vb{r},\omega')\right]=\mathcal{S}^{\alpha\beta}_{ij}(\omega)\delta\left(\omega-\omega'\right)\label{app1eq7}
\end{equation}
where, $\mathcal{S}^{\alpha\beta}_{ij}$ is an overlap matrix given by
\begin{equation}
\mathcal{S}_{ij}^{\alpha\beta}(\omega)=\frac{\hbar \omega^2}{\pi \epsilon_0 c^2}\frac{\vb{e}_\alpha\cdot \Im\left[\stackrel{\leftrightarrow}{\vb{G}}(\vb{r}_i,\vb{r}_j,\omega)\right]\cdot \vb{e}_\beta}{\mathcal{E}_{i,\alpha}(\omega)\mathcal{E}_{j,\beta}(\omega)}\label{app1eq8}
\end{equation}
where we used the Green's function identity $\sum_\lambda\int d^3\vb{s} \stackrel{\leftrightarrow}{\vb{G}}^{*T}_\lambda(\vb{r},\vb{s},\omega)\stackrel{\leftrightarrow}{\vb{G}}_\lambda(\vb{r}',\vb{s},\omega)=\frac{\hbar \omega^2}{\pi \epsilon_0 c^2} \Im\left[\stackrel{\leftrightarrow}{\vb{G}}(\vb{r},\vb{r}',\omega)\right]$ and the property $\stackrel{\leftrightarrow}{\vb{G}}_\lambda(\vb{r},\vb{r}',\omega)=\stackrel{\leftrightarrow}{\vb{G}}^{*T}_\lambda(\vb{r}',\vb{r},\omega)$.
The normalization factor $\mathcal{E}_{i,\alpha}(\omega)$ is obtained by imposing $\mathcal{S}_{ii}(\omega)=1$, to give
\begin{gather}
\mathcal{E}_{i,\alpha}(\omega)=\sqrt{\frac{\hbar \omega^2}{\pi \epsilon_0 c^2}\vb{e}_\alpha\cdot \Im\left[\stackrel{\leftrightarrow}{\vb{G}}(\vb{r}_i,\vb{r}_i,\omega)\right]\cdot \vb{e}_\alpha}\label{app1eq9}
\end{gather}
Since the coefficient functions $\vb{u}_{i,\alpha,\lambda}(\vb{r},\omega)$, are not always orthogonal. We must orthogonalize $\hat{b}_{i,\alpha}\left(\omega\right)$ into linearly independent modes by defining new emitter centered orthonormal modes as,
\begin{gather}
\hat{a}_{i,\alpha}(\omega)=\sum_{j=1}^N V_{ij}(\omega)\hat{b}_{i,\alpha}(\omega)\label{app1eq10},
\end{gather}
and vector amplitudes
\begin{gather}
\boldsymbol{\chi}_{i,\alpha,\lambda}(\vb{r},\omega)=\sum_{j=1}^N V_{ij}(\omega)\vb{u}_{j,\alpha,\lambda}(\vb{r},\omega)\label{app1eq11}
\end{gather}
where we can choose the transformation matrix $\mathbf{V}(\omega)$ by imposing the orthonormality condition,
\begin{equation}
[\hat{a}_{i,\alpha}(\omega),\hat{a}_{j,\beta}^\dagger(\omega')]=\sum_{i,j=1}^N \sum_{\alpha,\beta} V_{ij}(\omega) [\hat{b}_{i,\alpha}\left(\omega\right),\hat{b}_{j,\beta}^\dagger\left(\omega'\right)]V^*_{ji}(\omega')=\delta_{ij}\delta_{\alpha\beta}\delta(\omega-\omega')\label{app1eq12}
\end{equation}
The transformation matrix $\mathbf{V}(\omega)$ thus satisfies $\mathbf{V}(\omega)\mathbf{S}(\omega)\mathbf{V}^\dagger(\omega)=\mathbb{I}$. Eqs.~(\ref{app1eq10}) and (\ref{app1eq11}) also imply that $\hat{b}_{i,\alpha}(\omega)=\sum_{j=1}^N W_{ij}\hat{a}_{j,\alpha}(\omega)$ where, $\mathbf{W}(\omega)=\mathbf{V}(\omega)^{-1}$. Further, we can write the bosonic operator $ \hat{\vb{f}}_\lambda(\vb{r},\omega)$ in terms of the combination of linearly independent orthonormal set of operators, $\hat{a}_i(\omega)$, as
\begin{gather}
\hat{\vb{f}}_\lambda(\vb{r},\omega)=\sum_{i=1}^N\sum_\alpha\boldsymbol{\chi}^*_{i,\alpha,\lambda}(\vb{r},\omega) \hat{a}_{i,\alpha}(\omega)\label{app1eq13}
\end{gather}
We ignored the dark modes in the above expression as they do not interact with the quantum emitter and can be discarded as we formulate the Hamiltonians. For completeness, we indicate that if the dark modes are initially excited, including them may be required to accurately describe the state of the system. 
\par We use the definition of the fundamental bosonic modes in terms of the new dipole centered modes as per Eq.~(\ref{app1eq13}) and rewrite the field and interaction Hamiltonians as,
\begin{gather}
\hat{\mathcal{H}}_\text{f}=\int_0^\infty d\omega\hspace{2pt}\sum_{i=1}^N\sum_\alpha\hbar\omega\hat{a}^\dagger_{i,\alpha}(\omega)\hat{a}_{i,\alpha}(\omega) \label{app1eq14},
\end{gather}
and
\begin{gather}
 \hat{\mathcal{H}}_\text{int}=\int_0^\infty d\omega \sum_{i,j=1}^N\sum_\alpha g_{ij\alpha}(\omega) \hat{d}^{(+)}_{i,\alpha} \hat{a}_{j,\alpha}(\omega)+\text{H.c}\label{app1eq15}
\end{gather}
where, $g_{ij\alpha}(\omega)=\mathcal{E}_{i,\alpha}(\omega)W_{ij}(\omega)$. Thus we have a Hamiltonian description with only $N$ independent EM modes for each $\omega$. For single emitter ($N=1$) in an arbitrary orientation (no $\alpha$ dependence), Eqs.~(\ref{app1eq14}) and (\ref{app1eq15}) reduce to Eqs.~(\ref{eq:field H}) and (\ref{eq:light-matter H}) in the main text respectively with $g_{ii}(\omega)=g(\omega)$.

We can rewrite the electric field operator from Eq.~(\ref{app1eq3}) in terms of the orthonormal dipole centered modes, for a system of $N$ dipolar emitters in the near field, as $  \hat{\vb{E}}(\vb{r})= \hat{\vb{E}}^{(+)}(\vb{r})  +\text{H.c.}$, where
 \begin{gather}
 \hat{\vb{E}}^{(+)}(\vb{r})=\sum_{i=1}^N \sum_\alpha \int_0^\infty d\omega \vb{E}_{i,\alpha}(\vb{r},\omega) \hat{a}_{i,\alpha}(\omega)\label{app1eq17}
\end{gather}
for the mode function given by
\begin{equation} \label{eq:E vector}
\vb{E}_{i,\alpha}(\vb{r},\omega)=\frac{\hbar \omega^2}{\pi \epsilon_0 c^2}\sum_{j=1}^N \sum_{\beta} V^*_{ij}(\omega)\frac{\Im\left[\stackrel{\leftrightarrow}{\vb{G}}(\vb{r},\vb{r}_j,\omega)\right]}{\mathcal{E}_{j,\beta}(\omega)}\cdot\vb{e}_\beta
\end{equation} 

\section{Derivation of the Integro-Differential Equations for the Wavefunction Amplitudes}
\label{app:eom}

We derive the equations of motion for state amplitudes in the wave function $\lvert \psi(t)\rangle$ for a dipole coupled to a laser driven plasmonic cavity. The time evolution of the system can be described by the time dependent Schrödinger equation (TDSE),
\begin{gather}
i\hbar\frac{\partial \lvert \psi(t)\rangle}{\partial t}=\hat{\mathcal{H}} \lvert \psi(t)\rangle,\label{app2eq1}
\end{gather}
where $\hat{\mathcal{H}}$ is the total Hamiltonian. Upon acting the ansatz as per Eq.~[\ref{eq:wf ansatz}], we write,
\begin{equation}
\label{app2eq2}
\begin{split}
i\hbar\Bigg[ \Bigg.&\Dot{C}^{(g,0)}(t)\lvert g\rangle|\{0\}\rangle+\Dot{C}^{(e,0)}(t)\mathrm{e}^{-i\omega_{e} t}\lvert e\rangle|\{0\}\rangle+\int_0^\infty d\omega \hspace{2pt} \Dot{C}^{(g,1)}(\vb{r}_{qd},\omega,t) \mathrm{e}^{-i\omega t}\lvert g\rangle|\{\vb{1}(\vb{r}_{qd},\omega)\}\rangle\Bigg. \Bigg]\\=& \int_0^\infty d\omega\hspace{2pt}g(\omega)\Bigg[ \Bigg. \langle g\lvert d\rvert e\rangle^*C_{e0}(t)\mathrm{e}^{-i\omega_{e}t}|e\rangle|\{\vb{1}(\vb{r}_i,\omega)\}\rangle
     +\langle e\lvert d\rvert g\rangle C_{g1}(\vb{r}_{qd},\omega,t)\mathrm{e}^{-i\omega t} \lvert e\rangle|\{0\}\rangle\Bigg. \Bigg]\\
     &+ \int_0^\infty d\omega \hspace{2pt}\hbar D(\omega)\Bigg[ \Bigg.C_{g1}(\vb{r}_{qd},\omega,t)\lvert g\rangle|\{0\}\rangle+C_{g0}(t)\mathrm{e}^{-i\omega t}\lvert g\rangle\lvert\{\vb{1}(\vb{r}_{qd},\omega)\}\rangle\Bigg. \Bigg].
\end{split}
\end{equation}
In this full TDSE equation we have ignored the term corresponding to the $\lvert e\rangle\lvert\{\vb{1}(\vb{r}_{qd},\omega)\}\rangle$ state which is beyond the current wave function subspace. We proceed by projecting the complex conjugate of each state vectors $\langle g \rvert \langle\{0\}\rvert$, $\langle e \rvert \langle\{0\}\rvert$, and $\langle g \rvert \langle\{\vb{1}(\vb{r}_{qd},\omega)\}\rvert$ on the TDSE. The respective coupled equations thus obtained are
\begin{align}
i\hbar \dot{C}^{(g,0)}(t)&=\int_0^\infty d\omega\hspace{2pt} \hbar D(\omega)C_{g1}(\vb{r}_{qd},\omega,t),\label{app2eq3}\\
i\hbar\Dot{C}^{(e,0)}(t)\mathrm{e}^{-i\omega_{e} t}&=\int_0^\infty d\omega\hspace{2pt}g(\omega)\langle e\lvert d\rvert g\rangle C_{g1}(\vb{r}_{qd},\omega,t)\mathrm{e}^{-i\omega t}, \label{app2eq4}\\
i\hbar \Dot{C}^{(g,1)}(\vb{r}_{qd},\omega,t) \mathrm{e}^{-i\omega t}&=g(\omega) \langle g\lvert d\rvert e\rangle^* C_{e0}(t)\mathrm{e}^{-i\omega_{e}t}+\hbar D(\omega)C_{g0}(t)\mathrm{e}^{-i\omega t}. \label{app2eq5}
\end{align}
Now, formally integrating Eq.~[\ref{app2eq3}] gives us
\begin{gather}
C_{g0}(t)=-i\int_0^t dt' \int_0^\infty d\omega\hspace{2pt} D(\omega)C_{g1}(\vb{r}_{qd},\omega,t')+C_{g0}(0).\label{app2eq6}
\end{gather}
We rewrite Eq.~(\ref{app2eq5}) using the $C_{g0}(t)$ from the above equation as,
\begin{equation}
\label{app2eq7}
\begin{split}
i\hbar\Dot{C}^{(g,1)}(\vb{r}_{qd},\omega,t) =&g(\omega) \langle g\lvert d\rvert e\rangle^* C_{e0}(t)\mathrm{e}^{i(\omega-\omega_{e})t}+\hbar D(\omega)C_{g0}(0)\\&-i\hbar D(\omega)\int_0^t dt^\prime\int_0^\infty d\omega^\prime D(\omega^\prime) C_{g1}(\vb{r}_{qd},\omega^\prime,t').
\end{split}
\end{equation}
We change the dummy variables $t\rightarrow t'$ and $t'\rightarrow t''$ in Eq.~(\ref{app2eq7}) and formally integrate afterwards to obain,
 \begin{equation}
\label{app2eq8}
\begin{split}
i\hbar C_{g1}(\vb{r}_{qd},\omega,t) =&C_{g1}(\vb{r}_{qd},\omega,0)+\int_0^{t} dt^{\prime}\hspace{2pt}g(\omega) \langle g\lvert d\rvert e\rangle^* C_{e0}(t')\mathrm{e}^{i(\omega-\omega_{e})t'}+\hbar t D(\omega)C_{g0}(0)\\&-i\hbar D(\omega)\int_0^{t} dt^{\prime}\int_0^{t'} dt^{\prime\prime}\int_0^\infty d\omega^\prime D(\omega^\prime) C_{g1}(\vb{r}_{qd},\omega^\prime,t^{\prime\prime})
\end{split}
\end{equation}
By substituting the $C_{g1}(\vb{r}_{qd},\omega,t)$ in Eq.~(\ref{app2eq8}) to the Eq.~(\ref{app2eq4}), we obtain the equation of motion corresponding to the state amplitude $C_{e0}(t)$ as,
 \begin{equation}
\label{app2eq9}
\begin{split}
\Dot{C}^{(e,0)}(t)=&-\frac{1}{\hbar^2}\int_0^\infty d\omega\hspace{2pt}g(\omega)\hspace{2pt}\langle e\lvert d\rvert g\rangle\int_0^{t} dt^{\prime}\hspace{2pt}g(\omega) \langle g\lvert d\rvert e\rangle^* C_{e0}(t')\mathrm{e}^{i(\omega-\omega_{e})t'}\mathrm{e}^{-i(\omega-\omega_e) t}\\& -\frac{1}{\hbar}\int_0^\infty d\omega\hspace{2pt}g(\omega)\hspace{2pt}\langle e\lvert d\rvert g\rangle t D(\omega) C_{g0}(0)\mathrm{e}^{-i(\omega-\omega_e) t}-\frac{1}{\hbar^2}\int_0^\infty d\omega \hspace{2pt} g(\omega) \langle e\lvert d \rvert g\rangle C_{g1}(\vb{r}_{qd},\omega,0)\mathrm{e}^{-i(\omega-\omega_{e})t}\\
&+\frac{i}{\hbar}\int_0^\infty d\omega\hspace{2pt}g(\omega)\langle e\lvert d\rvert g\rangle D(\omega)\int_0^{t} dt^{\prime}\int_0^{t'} dt^{\prime\prime}\int_0^\infty d\omega^\prime D(\omega^\prime) C_{g1}(\vb{r}_{qd},\omega^\prime,t^{\prime\prime})\mathrm{e}^{-i(\omega-\omega_e) t}
\end{split}
\end{equation}
We make an assumption to treat the system of equations which are linear to the laser field amplitude $D(\omega)$. So we ignore the last term of the previous equation to write,
 \begin{equation}
\label{app2eq10}
\begin{split}
\Dot{C}^{(e,0)}(t)=&-\frac{1}{\hbar^2}\int_0^{t} dt^{\prime}\int_0^\infty d\omega\hspace{2pt}\lvert g(\omega)\rvert^2\hspace{2pt}\lvert \langle e\lvert d\rvert g\rangle\rvert^2  C_{e0}(t')\mathrm{e}^{-i(\omega-\omega_{e})(t-t')}\\& -\frac{t}{\hbar}\int_0^\infty d\omega\hspace{2pt}g(\omega)\hspace{2pt}\langle e\lvert d\rvert g\rangle  D(\omega) C_{g0}(0)\mathrm{e}^{-i(\omega-\omega_e) t}\\&-\frac{1}{\hbar^2}\int_0^\infty d\omega\hspace{2pt}g(\omega)\hspace{2pt}\langle e\lvert d\rvert g\rangle C_{g1}(\vb{r}_{qd},\omega,0) \mathrm{e}^{-i(\omega-\omega_e) t}
\end{split}
\end{equation}
Hence we write,
\begin{equation}
\Dot{C}^{(e,0)}(t)=\int_0^{t} dt^{\prime}\mathcal{K}^{(e,0)}(t-t')C_{e0}(t')+S(t) \label{app2eq11}
\end{equation}
where,
\begin{equation}
\mathcal{K}(t-t')= -\int_0^\infty d\omega \underbrace{\frac{1}{\hbar^2} \lvert g(\omega)\rvert^2\hspace{2pt}\lvert \langle e\lvert d\rvert g\rangle\rvert^2}_{\mathcal{K}(\omega)} \mathrm{e}^{-i(\omega-\omega_{e})(t-t')}\label{app2eq12}
\end{equation}
and,
\begin{align}
\mathcal{S}(t)=&-\frac{1}{\hbar}\int_0^\infty d\omega\hspace{2pt} \sqrt{\mathcal{K}(\omega)}  C_{g1}(\vb{r}_{i},\omega,0)\mathrm{e}^{-i(\omega-\omega_e) t} \label{app2eq13}\\
&-t\int_0^\infty d\omega\hspace{2pt} \sqrt{\mathcal{K}(\omega)}  D(\omega) C_{g0}(0)\mathrm{e}^{-i(\omega-\omega_e) t},\nonumber
\end{align}
The remaining two state amplitudes can be obtained from Eqs.~(\ref{app2eq6}) and (\ref{app2eq8}) as,
\begin{align}
C_{g0}(t)&=-i\int_0^t dt' \int_0^\infty d\omega\hspace{2pt} D(\omega)C_{g1}(\vb{r}_{qd},\omega,t')+C_{g0}(0),\label{app2eq14}\\
C_{g1}(\vb{r}_{qd},\omega,t)&=-i\int_0^{t} dt^{\prime}\hspace{2pt}\sqrt{\mathcal{K}(\omega)}C_{e0}(t')\mathrm{e}^{i(\omega-\omega_{e})t'}-i t D(\omega)C_{g0}(0). \label{app2eq15}
\end{align}

\subsection{Electric Field Amplitude and Intensity}

 The Wigner-Weisskopf wavefunction ansatz in Eq.~[\ref{eq:wf ansatz}] can be employed to calculate the expectation value of the electric field operator from Eq.~(\ref{app1eq17}). For a single emitter, we write,
 \begin{align} 
\centering
\hat{\vb{E}}^{(+)}(\vb{r}) |\psi(t)\rangle &=\int_0^\infty d\omega \vb{E}(\vb{r},\omega) \hat{a}(\omega) \int_0^\infty d\omega' \int d^3 \vb{r}' C_{g1}(\vb{r}',\omega',t)\mathrm{e}^{-i\omega' t} |g\rangle|\{\vb{1}(\vb{r}',\omega')\}\rangle\nonumber\\
&=\int_0^\infty d\omega \vb{E}(\vb{r},\omega) C_{g1}(\vb{r},\omega,t)\mathrm{e}^{-i\omega t} |g\rangle |\{0\}\rangle  \label{eq E+ derive}
\end{align}
So, the expectation value of electric field amplitude can be written as,
\begin{gather}
\langle  \hat{\vb{E}}^{(+)}(\vb{r},t)\rangle=\langle \psi (t) | \hat{\vb{E}}^{(+)}(\vb{r}) |\psi(t)\rangle  \label{app2eq17}\\\nonumber\\
\langle  \hat{\vb{E}}^{(+)}(\vb{r},t)\rangle=  \int_0^\infty d\omega \vb{E}(\vb{r},\omega) C^*_{g0}(t)C_{g1}(\vb{r},\omega,t)\mathrm{e}^{-i\omega t}\label{eq E amp}
\end{gather}
We proceed to calculate the expectation value of intensity of electric field, $I(\vb{r},t)$. We continue from the Eq.~(\ref{eq E+ derive}) to write,
\begin{align} 
\centering
\hat{\vb{E}}^{(-)}(\vb{r})\hat{\vb{E}}^{(+)}(\vb{r}) |\psi(t)\rangle &= \hat{\vb{E}}^{(-)}(\vb{r})  \int_0^\infty d\omega^\prime \vb{E}(\vb{r},\omega^\prime) C_{g1}(\vb{r},\omega^\prime ,t)\mathrm{e}^{-i\omega^\prime t} |g\rangle |\{0\}\rangle \nonumber\\
&= \int_0^\infty d\omega \vb{E}(\vb{r},\omega) \hat{a}^\dagger(\omega) \int_0^\infty d\omega^\prime \vb{E}(\vb{r},\omega^\prime) C_{g1}(\vb{r},\omega^\prime ,t)\mathrm{e}^{-i\omega^\prime t} |g\rangle |\{0\}\rangle \nonumber\\
&=  \int_0^\infty d\omega \lvert \vb{E}(\vb{r},\omega) \rvert^2   C_{g1}(\vb{r},\omega ,t)\mathrm{e}^{-i\omega t} \lvert g\rangle \lvert \{\vb{1}(\omega)\}\rangle
\end{align}
The expectation value is
\begin{align} 
\centering
\langle\psi(t)\rvert \hat{\vb{E}}^{(-)}(\vb{r})\hat{\vb{E}}^{(+)}(\vb{r}) \lvert \psi(t)\rangle &= \int_0^\infty d\omega C^*_{g1}(\vb{r},\omega,t)e^{i\omega t}\langle g\rvert \langle \vb{1}(\omega)\rvert  \int_0^\infty d\omega^\prime \lvert \vb{E}(\vb{r},\omega^\prime) \rvert^2   C_{g1}(\vb{r},\omega^\prime ,t)\mathrm{e}^{-i\omega^\prime t} \lvert g\rangle \lvert \{\vb{1}(\omega^\prime)\}\rangle \nonumber\\
&=  \int_0^\infty d\omega  \lvert \vb{E}(\vb{r},\omega) \rvert^2  \lvert C_{g1}(\omega,t)\rvert^2 .\label{eq Intensity derive}
\end{align}
Hence, the expression for electric field intensity is,
\begin{equation}
I(\vb{r},t)=\langle  \hat{\vb{E}}^{(-)}(\vb{r})\hat{\vb{E}}^{(+)}(\vb{r}) \rangle =  \int_0^\infty d\omega  \lvert \vb{E}(\vb{r},\omega) \rvert^2  \lvert C_{g1}(\omega,t)\rvert^2 \label{eq Intensity}
\end{equation}
where,
\begin{gather}
\vb{E}(\vb{r},\omega)=\frac{\hbar \omega^2}{\pi \epsilon_0 c^2}\frac{\Im\left[\stackrel{\leftrightarrow}{\vb{G}}(\vb{r},\vb{r},\omega)\right]}{g(\omega)}\cdot\vb{e}_z .\label{eq E vector single}
\end{gather}

 \section{IDE solution via Laplace transforms}
 \label{app:Laplace IDE}
 
We solve the integro-differential Eq.~(\ref{app2eq11}) for $C_{e0}(t)$ using Laplace transform technique. Assuming the spectral density $\mathcal{K}(\omega)$ takes the peak shape, we model it as combination of Lorentzian line shapes. We write, 
\begin{equation}
\mathcal{K}(\omega)=\sum_{j=1}^n \frac{A_j}{\pi} \frac{B_j}{(\omega-\Omega_j)^2+B_j^2}  \label{app2eq16}
\end{equation}
where $A_j$s, $B_j$s and $\Omega_j$s are the corresponding Lorentzian fitting parameters and $n$ is the number of  such Lorentzians necessary to fit the $\mathcal{K(\omega)}$. We use this Lorentzian $\mathcal{K}(\omega)$in the Eq.~(\ref{app2eq12}) and the following the Fourier integral gives the delay Kernel,
\begin{equation}
\mathcal{K}(\tau)=\sum_{j=1}^n A_j\mathrm{e}^{-B_j \tau-i(\Omega_j-\omega_e)\tau}\label{app2eq17}
\end{equation}
where we take $t-t'=\tau$. Although the approach can be employed seamlessly for any sum of Lorentzians, for the MNP in the study, $n=3$ is typically the number of Lorentzians sufficient for fitting the spectral density peaks. The fit parameters with three Lorentzians for the $\mathcal{K}(\omega)$ nanosphere-QD system in discussion is given in Table \ref{tab:fitparams}.
\begin{table}[h]
\caption{The Lorentzian fitting parameters for $\mathcal{K(\omega)}$ for the sphere-QD separation distance $h=2$\,nm (Fig: \ref{fig:kernels}a) and $h=10$\,nm (Fig: \ref{fig:kernels}b).}
\label{tab:fitparams}
\begin{ruledtabular}
\begin{tabular}{cccccccccc}
 & $A_1$ (THz$^2$) & $A_2$ (THz$^2$) & $A_3$ (THz$^2$) & $B_1$ (eV) & $B_2$ (eV) & $B_3$ (eV) & $\Omega_1$ (eV) & $\Omega_2$ (eV) & $\Omega_3$ (eV) \\
\hline
$h=2$\,nm & 2.1204 & 102.307 & 175.1694 & 0.0243& 0.03615 &  0.02507 & 2.757 & 2.9498 & 2.972 \\
$h=10$\,nm & 0.5759 & 0.47819 & 0.9852 & 0.03212 & 0.02622 & 0.03421 & 2.757 & 2.8812 & 2.932 \\
\end{tabular}
\end{ruledtabular}
\end{table}\\

 For $n=3$, considering $C_{e0}(t)=y(t)$, we write,
\begin{equation}
	y'(t)=\int_{0}^{t}\left[A_1{\rm e}^{-B_1\tau-i\left(\Omega_1-\omega_1\right)\tau}+A_2 {\rm e}^{-B_2\tau-i\left(\Omega_2-\omega_1\right)\tau}+A_3 {\rm e}^{-B_3\tau-i\left(\Omega_3-\omega_1\right)\tau}\right]y(\tau)d\tau + \mathcal{S}(t) \label{app2eq18}
\end{equation}
The initial condition is defined by $y(0) = y_0$, an arbitrary. Let's call $B_j+i\left(\Omega_j-\omega_1\right)=\Delta_j$,
\begin{equation}
	y'(t)=\int_{0}^{t}\left[A_1{\rm e}^{-\Delta_1\tau}+A_2 {\rm e}^{-\Delta_2\tau}+A_3 {\rm e}^{-\Delta_3\tau}\right]y(\tau)d\tau +\mathcal{S}(t) \label{app2eq19}
\end{equation}
Using the convolution property and taking the Laplace transform of the Eq.~(\ref{app2eq19}), we get
\begin{align}
	s\mathcal{L}\{y(t)\}-y(0)&=\mathcal{L}\{A_1{\rm e}^{-\Delta_1\tau}+A_2 {\rm e}^{-\Delta_2\tau}+A_3 {\rm e}^{-\Delta_3\tau}\}\mathcal{L}\{y(t)\}+\mathcal{L}\{\mathcal{S}(t)\},  \label{app2eq20}\\
	\implies sY(s)-y_0&=\left(\frac{A_1}{s+\Delta_1}+\frac{A_2}{s+\Delta_2}+\frac{A_3}{s+\Delta_3}\right)Y(s)+\mathcal{S}(s),  \label{app2eq21}
\end{align}
where, $\mathcal{L}\{y(t)\}=Y(s)$ and $\mathcal{L}\{\mathcal{S}(t)\}=\mathcal{S}(s)$.
We can write this as,
\begin{eqnarray}
	Y(s)=\mathcal{S}(s)G(s)+y_0 G(s)\label{app2eq22}
\end{eqnarray}
where the rational polynomial $G(s)$ is given by
\begin{eqnarray}
	G(s)=\frac{(s+\Delta_1)(s+\Delta_2)(s+\Delta_3)}{s(s+\Delta_1)(s+\Delta_2)(s+\Delta_3)-A_1(s+\Delta_2)(s+\Delta_3)-A_2(s+\Delta_1)(s+\Delta_3)-A_3(s+\Delta_1)(s+\Delta_2)}\label{app2eq23}
\end{eqnarray}
Similarly, we can also analytically calculate the Laplace transform $\mathcal{S}(s)$, which is another rational polynomial in $s$, from Eq.~(\ref{app2eq13}). Here we assume Lorentzian lineshape for the $\mathcal{K}(\omega)$ dominated products $ \sqrt{\mathcal{K}(\omega)}  C_{g1}(\vb{r}_{i},\omega,0)$ and $ \sqrt{\mathcal{K}(\omega)}  D(\omega) C_{g0}(0)$.  Hence we rewrite Eq.~(\ref{app2eq22}) as a total rational polynomial,
\begin{equation}
Y(s)=\frac{Q(s)}{P(s)}.\label{app2eq24}
\end{equation}
So we can write the final solution, the inverse Laplace transform \cite{Polyanin2008handbook} as,
\begin{equation}
y(t)=\sum_{k=1}^n \sum_{l=1}^{m_k}\frac{\Phi_{kl}(\lambda_k)}{(m_k-l)! (l-1)!} t^{m_k-l}\text{exp}(\lambda_k t), \label{a2eq:25}
\end{equation}
where $\lambda_k$ are the roots of the polynomial $P(s)$ and $\Phi_{kl}(s)=\frac{d^{l-1}}{ds^{l-1}}\left[\frac{Q(s)}{H_k(s)}\right]$ and $H_k(s)=\frac{P(s)}{(s-\lambda_k)^{m_k}}$.

\subsection{Analytical solution for a single Lorentzian kernel}
We begin with a single Lorentzian spectral kernel function,
\begin{equation}
\mathcal{K}(\omega) = \frac{A}{\pi}\,\frac{B}{(\omega-\Omega)^2 + B^2}.
\label{eq:K_omega}
\end{equation}
Let the qubit transition frequency be $\omega_e$. Working in the interaction frame at frequency \(\omega_e\), time-domain memory kernel that corresponds to the Lorentzian of Eq.~\eqref{eq:K_omega} is obtained by the inverse Fourier transform. The time-domain kernel takes the exponential form
\begin{equation}
\mathcal{K}(\tau) = A\,{\rm e}^{-B\tau} {\rm e}^{-i(\Omega-\omega_e) \tau} = A\,{\rm e}^{-\tilde{B}\,\tau},\qquad \tau\ge0,
\label{eq:memory_kernel_time}
\end{equation}
where we define the complex parameter $\tilde{B} = B + i (\Omega-\omega_e)$.

Now consider the integro-differential equation for $C_{e0}(t)$ with no external source ($S(t)=0$),
\begin{equation}
\dot{C}_{e0}(t) = -\int_0^t \mathcal{K}(\tau)\,C_{e0}(t-\tau)\,d\tau.
\label{eq:IDE_conv}
\end{equation}
where also we set the initial condition $C_{e0}(0)=1$.

Taking the Laplace transform, we have
\begin{align}
s\,C_{e0}(s)-1&= -\frac{A}{s+\tilde{B}}\,C_{e0}(s),\\
C_{e0}(s) &= \frac{s + \tilde{B}}{s^2 + \tilde{B}s + A}.
\label{eq:C_s}
\end{align}
where we used the convolution theorem.

We proceed expand the Laplace domain solution using completing the square method to write,
\begin{equation}
C_{e0}(s) = \frac{s + 2\alpha}{(s+\alpha)^2 + b^2}
= \frac{s+\alpha}{(s+\alpha)^2 + b^2} + \frac{\alpha}{(s+\alpha)^2 + b^2},
\end{equation}
where, we define $2\alpha=\tilde{B}$ and $2b=\sqrt{4A-\tilde B^2}$

The inverse Laplace transform of $C_{e0}(s)$ gives the time-domain solution for the excited state amplitude which is shown in Eq.~(\ref{eq:Ce0 analytical}) of the main text,
\begin{equation}
C_{e0}(t) =  {\rm e}^{-\tilde Bt/2}\left[\cos(bt)+\frac{\tilde B}{2b}\sin(bt)\right].
\label{eq:Ce0_solution}
\end{equation}

Now, we have the photon amplitude as,
\begin{equation}
C_{g1}(\omega,t) = -i \int_0^t C_{e0}(t') \sqrt{\mathcal{K}(\omega)} \, 
e^{i \delta t'} \, \mathrm{d}t',
\label{eq:Cg1integral_delta}
\end{equation}
where the detuning is $\delta = \omega - \omega_e$.

We employ Eq.~(\ref{eq:Ce0_solution}) to write,
\begin{equation}
C_{g1}(\omega,t) = -i \sqrt{\mathcal{K}(\omega)} \int_0^t e^{-\frac{\tilde{B}}{2} t'} 
\left[ \cos(bt') + \frac{\tilde{B}}{2b} \sin(bt') \right] e^{i \delta t'} \, \mathrm{d}t'.
\end{equation}

Using the standard integrals, we obtain,
\begin{equation}
C_{g1}(\omega,t) = -i\,\frac{\sqrt{\mathcal{K}(\omega)}}{(-\tfrac{\tilde{B}}{2}+i\delta)^2+b^2} \left[ e^{(-\tfrac{\tilde{B}}{2}+i\delta)t}\!\left[(i\delta-\tilde{B})\cos(bt)+\Big(b+\tfrac{\tilde{B}}{2b}(i\delta-\tfrac{\tilde{B}}{2})\Big)\sin(bt)\right] +(-i\delta+\tilde{B}) \right].
\end{equation}

We can rewrite the photon amplitude in terms of the detuning $\delta$ as
\begin{equation}
C_{g1}(\delta,t) = -i \frac{\sqrt{\mathcal{K}(\delta)}}{A - \delta^2 - i B \delta} \Big[ e^{(-\tfrac{\tilde{B}}{2}+i\delta) t} \big( (i \delta - \tilde{B}) \cos(bt) + \big(b + \frac{i \delta \tilde{B}}{2b} - \frac{\tilde{B}^2}{4b}\big) \sin(bt) \big) + (\tilde{B} - i \delta) \Big]. \label{eq:Cg1_full_detuning}
\end{equation}
where $\mathcal{K}(\delta) = \mathcal{K}(\omega_e + \delta)$ is the spectral density evaluated at $\delta$.

In the long-time limit $t \to \infty$, the exponentially decaying term vanishes and only the constant term remains. Setting $\Omega = \omega_e$ so that $\tilde{B} = B$, we obtain
\begin{equation}
C_{g1}(\delta, \infty) = -i \frac{\sqrt{\mathcal{K}(\delta)}\, (B - i \delta)}{A - \delta^2 - i B \delta}. \label{eq:Cg1_long_time} 
\end{equation}

It is convenient to express the full time dependent photon amplitude \(C_{g1}(\delta,t)\) in terms of its long-time limit,
\begin{equation}
C_{g1}(\delta,t) = C_{g1}(\delta, \infty) 
\Bigg[ 1 - e^{(i \delta - B/2) t} 
\Bigg( \cos(bt) + \frac{b + \frac{i \delta B}{2b} - \frac{B^2}{4b}}{i \delta - B} \, \sin(bt) \Bigg) \Bigg]. \label{eq:Cg1_full_long_time} 
\end{equation}

This expression becomes Eq.~(\ref{eq:Cg1 analytical}) in the main text when we approximate to terms linear to $b$ as,
\begin{equation}
C_{g1}(\delta,t) = C_{g1}(\delta, \infty) 
\Bigg[ 1 - e^{(i \delta - B/2) t} 
\Bigg( \cos(bt) - p(\delta) \, \sin(bt) \Bigg) \Bigg]. \label{eq:Cg1_full_long_time} 
\end{equation}
where, $p(\delta)=b/(B-i \delta)$.

\section{Benchmarking the Lorentzian Kernel Model Against Established Results} 
\label{app:comparison}
To validate our methodology, we benchmark our results against published studies of plasmon-emitter interactions in experimentally realized cavity structures. While complex cavity geometries typically require numerical simulations to compute spectral densities, our pseudo-mode approximation enables analytical treatment of these numerically obtained results.

\begin{figure}[h]
    \centering
    \includegraphics[width=0.6\linewidth]{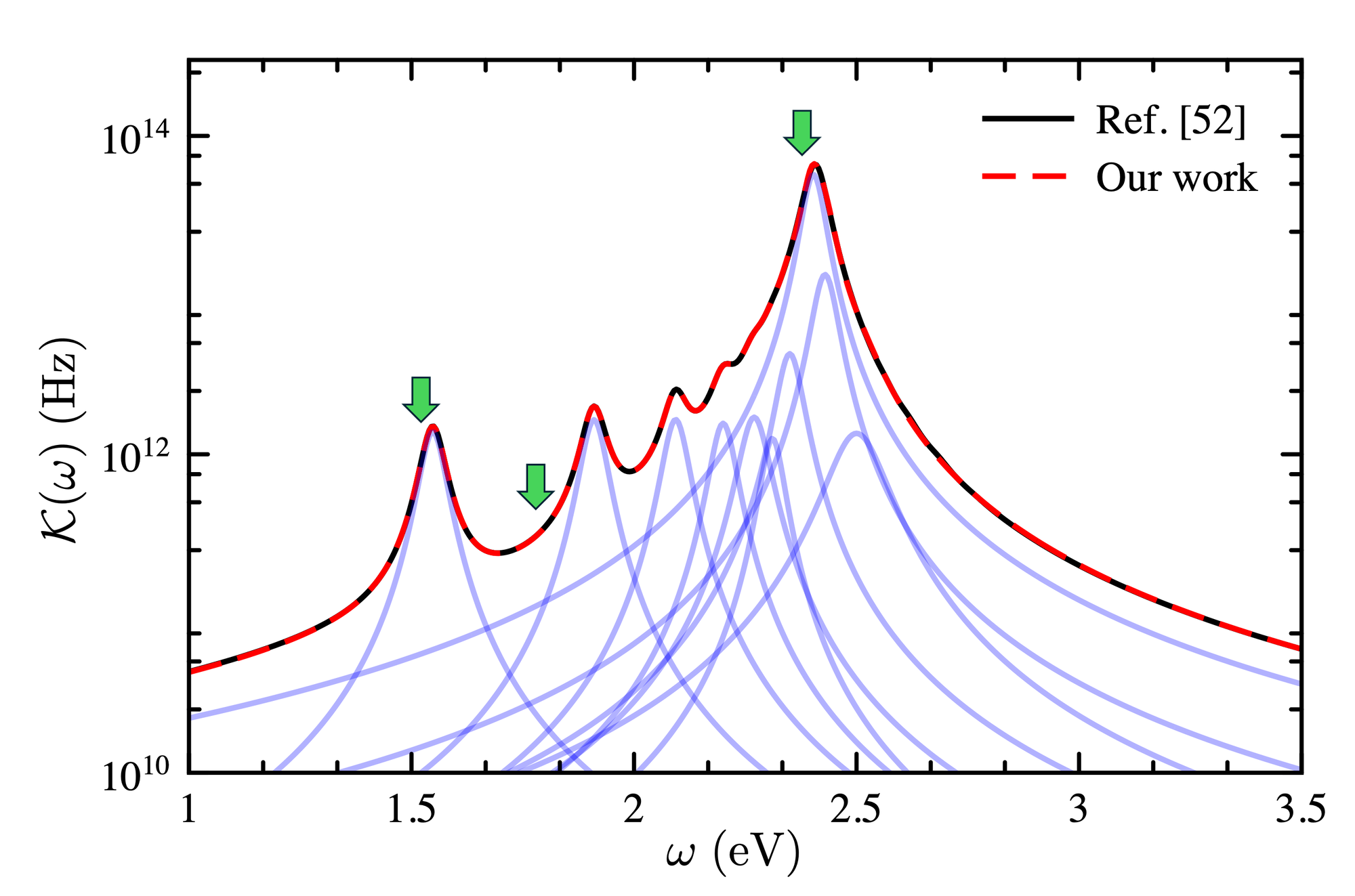}
    \caption{
        Spectral density $\mathcal{K}(\omega)$ (in Hz) of a dipolar qubit in a nanoparticle-on-mirror (NPoM) cavity. The exact result from~\cite{CuarteroGonzalez2020} (black solid line) is compared with its multi-Lorentzian fit (red dashed line), with individual Lorentzian components shown as faded blue solid lines. Green arrows indicate the qubit frequencies $\omega_e$ at which population dynamics are analyzed in Fig.~\ref{prb_comparison}.}
    \label{kernel_NPoM}
\end{figure}

We specifically compare with the reported dynamics of a dipolar exciton in a nanoparticle-on-mirror (NPoM) cavity under spontaneous emission conditions~\cite{CuarteroGonzalez2020}. The spectral density $J_\mu(\omega)$ for a dipolar exciton can be directly related to our $\mathcal{K}(\omega)$ through
\begin{equation}
    \mathcal{K}(\omega) = \frac{\sqrt{\epsilon_b}}{8} J_\mu(\omega),
    \label{a3eq:1}
\end{equation}
where we use $\epsilon_b=4$ is the permittivity of the surrounding dielectric medium from the reference. Following the procedure outlined in Section~\ref{sec:LorentzianFit}, we approximate the numerically computed $\mathcal{K}(\omega)$ for the NPoM cavity using a sum of Lorentzian distributions. As shown in Fig.~\ref{kernel_NPoM}, a superposition of 10 Lorentzians provides an excellent fit to the spectral density. The corresponding fit parameters are listed in Table.~\ref{tab:NPoM fit}.

\begin{table}[h]
\caption{The 10 Lorentzian fitting parameters for $\mathcal{K}(\omega)$ for the NPoM cavity spectral density data from Ref.~\cite{CuarteroGonzalez2020}.}
\label{tab:NPoM fit}
\begin{ruledtabular}
\begin{tabular}{cccc}
 Peak  Number& $A_i$ (THz$^2$) & $B_i$ (eV) & $\Omega_i$ (eV) \\
\hline
 1  & 31.8917   & 0.0303 & 1.5478 \\
 2  & 37.9805   & 0.0304 & 1.9101 \\
 3  & 38.1770   & 0.0304 & 2.0936 \\
 4  & 35.7083   & 0.0300 & 2.2003 \\
  5  & 45.2399   & 0.0348 & 2.2700 \\
 6  & 25.9778   & 0.0272 & 2.3110 \\
 7  & 99.6008   & 0.0306 & 2.3500 \\
 8  & 1196.2133 & 0.0276 & 2.4032 \\
 9 & 304.6571  & 0.0298 & 2.4292 \\
 10  & 67.2896   & 0.0654 & 2.5000 \\
\end{tabular}
\end{ruledtabular}
\end{table}

\begin{figure}[h]
    \centering
    \includegraphics[width=\linewidth]{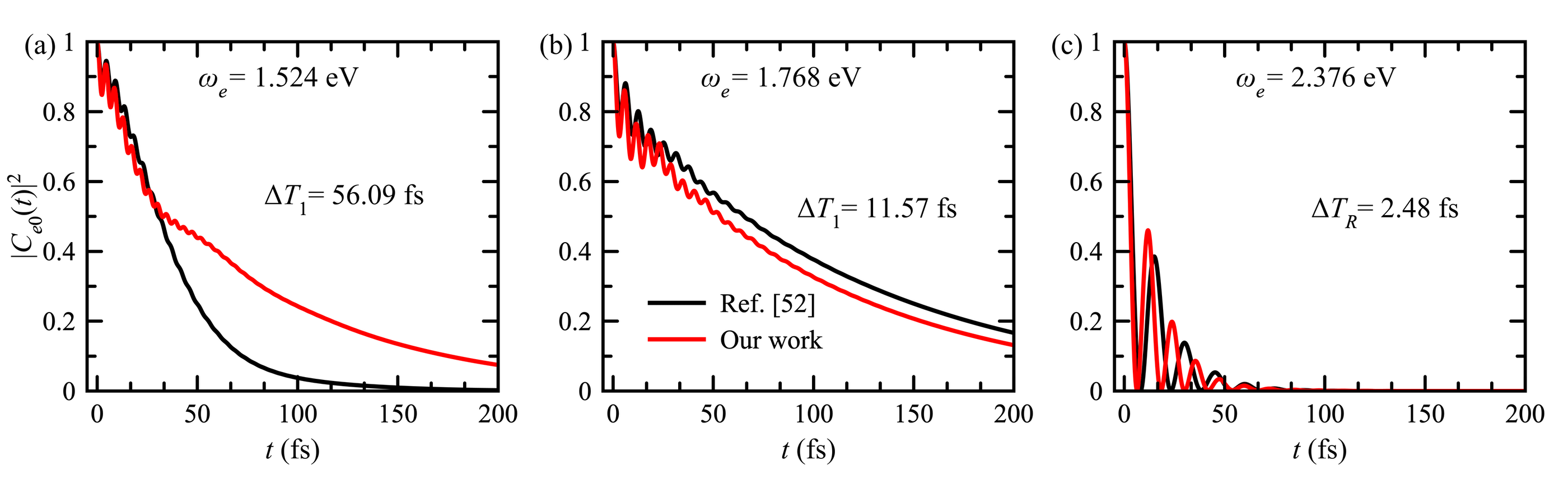}
    \caption{Excited-state population dynamics $|C_{e0}(t)|^2$ at three representative frequencies: (a) 1.52 eV, (b) 1.78 eV, and (c) 2.376 eV. Comparison between reference results (blue) from Ref.~\cite{CuarteroGonzalez2020} and our Lorentzian-approximated kernel calculations (red). All panels share the population scale shown in (a). Numerical differences in decay times ($\Delta T_1$) and Rabi periods ($\Delta T_R$) are annotated in each panel.}
    \label{prb_comparison}
\end{figure}

This pseudo-mode representation enables efficient computation of exciton population dynamics via Eq.~(\ref{eq:Ce0}). Figure~\ref{prb_comparison} compares the population dynamics at three representative frequencies along the spectral density. Using Laplace transform techniques, we compute time traces of the population dynamics at the frequencies marked in Fig.~\ref{kernel_NPoM} (green arrows). While our approach captures all regime-specific trends, we observe some quantitative differences in temporal resolution.

Near the dipolar plasmonic resonance at 1.52~eV (Fig.~\ref{prb_comparison}a), we observe the most significant difference in the decay dynamics between our model and the exact numerical result, with our model exhibiting a longer decay time ($\Delta T_1 \approx 56$~fs). At higher energies, specifically at $\omega_e = 1.78$~eV (in the post-resonance dip region), the Markovian decay profiles show closer agreement (Fig.~\ref{prb_comparison}b), with a reduced discrepancy of $\Delta T_1 \approx 11.57$~fs. In the strong-coupling regime near the 2.376~eV plasmonic peak, our model successfully captures the non-Markovian features, including Rabi oscillations (Fig.~\ref{prb_comparison}c), with only a 2.48~fs difference in the oscillation period. The aforementioned findings are robust even when using a spectral fit for $\mathcal{K}(\omega)$ with fewer Lorentzians and slightly relaxed quality criteria.

 \end{widetext}

\end{document}